\PassOptionsToPackage{hyphens}{url} 
\documentclass[bibyear]{aa}  

\usepackage{graphicx}
\usepackage{txfonts}

\usepackage{amsmath,amssymb}  
\usepackage{siunitx}          
\usepackage{upgreek}          
\usepackage{bm}               
\usepackage{booktabs}         
\usepackage{tabularx}         
\usepackage[dvipsnames]{xcolor}           
\usepackage{multirow}
\usepackage{adjustbox}
\usepackage[switch]{lineno}

\usepackage[
    colorlinks=true,    
    citecolor=blue,     
    linkcolor=blue,     
    urlcolor=blue,      
    breaklinks=true     
]{hyperref}

\usepackage[authoryear]{natbib}
\bibpunct{(}{)}{;}{a}{}{,}
\defcitealias{hovatta19}{H19}

\DeclareMathOperator{\sinc}{sinc}
\newcommand*\rad{~rad\,m$^{-2}$}

\newcommand{\tclean}{\texttt{tclean}}

\begin{document} 

   \title{Multiband ALMA polarimetry of the jet base and kiloparsec-scale jet of 3C 273: A multicomponent Faraday structure in the nucleus}

   \author{T. Hovatta\inst{1,2,3}
   \thanks{\href{mailto:talvikki.hovatta@aalto.fi}{talvikki.hovatta@aalto.fi}} \orcid{0000-0002-2024-8199}, 
          S.~P.~O'Sullivan\inst{4}\orcid{0000-0002-3968-3051}, 
          I. Mart\'i-Vidal\inst{5,6}\orcid{0000-0003-3708-9611}, 
          T.~Savolainen\inst{2,3,7}\orcid{0000-0001-6214-1085}
          }

   \institute{Finnish Centre for Astronomy with ESO (FINCA), University of Turku, FI-20014 Turku,  Finland
              \and
              Aalto University Mets\"ahovi Radio Observatory, Mets\"ahovintie 114, FI-02540 Kylm\"al\"a, Finland
              \and
              Aalto University Department of Electronics and Nanoengineering, PL~15500, FI-00076 Aalto, Finland
              \and 
              Departamento de Física de la Tierra y Astrofísica \& IPARCOS-UCM, Universidad Complutense de Madrid, 28040 Madrid, Spain
              \and
              Dpt. Astronomia i Astrofísica, Universitat de València, C/ Dr. Moliner 50, 46120 Burjassot, Spain
              \and
              Observatori Astronòmic, Universitat de València, C/ Cat. José Beltrán 2, 46980 Paterna, Spain
              \and
              Max-Planck-Institut f\"ur Radioastronomie, Auf dem H\"ugel 69, DE-53121 Bonn, Germany}

   \date{Received 11 May 2026 / Accepted 06 July 2026}

 
  \abstract
   {}
   {
   Polarization observations at millimeter wavelengths can be used to study magnetized plasma in jets launched by supermassive black holes. We used multiband Atacama Large Millimeter Array (ALMA) data to study the polarization structure of both the nucleus and the kiloparsec-scale jet of the archetypal quasar 3C 273.
   }
   {
   We modeled the wavelength-dependent polarization of the nucleus using multiband observations (2, 1.3, and 0.85 mm) and applied QU-fitting to constrain Faraday rotation models with one or more polarized components. We also produced total intensity and polarization maps of the kiloparsec-scale jet to study the evolution of the magnetic field structure along and across the jet.
   }
   {
   The data of the nucleus were best fit by two Faraday components: a high rotation measure (RM) component with ${\rm RM} = +(2.6 \pm 0.1) \times 10^{5}~\si{rad.m^{-2}}$ and comparable RM dispersion, and a second component with ${\rm RM} = +(1.5 \pm 0.2) \times 10^{4}~\si{rad.m^{-2}}$ and lower dispersion. The high RM is comparable to previous ALMA observations at 1.3 mm, although we found clear evidence of variability when compared to previous single-band ALMA studies, highlighting the importance of broad wavelength coverage. On kiloparsec scales, the millimeter polarization structure closely resembled that observed at centimeter wavelengths, revealing a complex magnetic field configuration around the jet head and brightest hotspot region.
   }
   {
   Our results indicate a multicomponent, time-variable Faraday screen in the nuclear region of 3C 273 that is likely associated with a dense, magnetized environment close to the jet base. Future spatially resolved millimeter polarimetry with the Event Horizon Telescope will be crucial to disentangling these components and directly localizing the high-RM emission region.
   }

   \keywords{polarization -- galaxies: magnetic fields --
                quasars: individual: 3C~273 --
                galaxies: jets
               }

    \titlerunning{ALMA multiband polarization observations of 3C~273}

   \maketitle
%
\section{Introduction}
 The archetypal quasar 3C~273 was the first radio source identified to be at a high redshift $z=0.158$ due to its optical spectrum \citep{schmidt1963}; this subsequently made it the first quasar discovered. Its optical images showed a bright star along with a ``faint wisp or jet'' \citep{schmidt1963}, which were coincident with the positions derived from the first radio observations of the nucleus and the jet by lunar occultation measurements that also revealed the flat spectrum of the nucleus \citep{hazard63}. The kiloparsec-scale jet of 3C~273 has since been extensively studied in the radio (see, e.g., \citealt{conway93} for a comprehensive discussion of the early radio observations) and other wavelengths, from the infrared to X-rays \citep[e.g.,][]{willingale81,meisenheimer86, meisenheimer89, conway94, bahcall95, neumann97, roeser00, sambruna01, jester01, uchiyama06, jester07,marchenko17}. 
A complete set of radio observations from the Very Large Array (VLA) at multiple radio frequencies was published in \cite{perley17}. Recently, the first millimeter-wavelength total intensity images of the kiloparsec-scale jet of 3C~273 taken by the Atacama Large Millimeter Array (ALMA) were published by \cite{komugi22}. 

The kiloparsec-scale jet of 3C~273 consists of multiple knots linked by diffuse emission. The first knot, A, is about 13\arcsec{} from the nucleus and is followed by subsequent knots into the jet head region, which itself contains another three knots, H3, H2, and H1 \citep[listed in the order toward the head; e.g.,][]{jester05}. At radio wavelengths, knot H2 dominates the emission, whereas at X-ray energies knot A dominates and the downstream knots fade away toward the jet head \citep[e.g.,][]{sambruna01}. In the optical, the jet is bright at both knots A and H3; the downstream knots H2 and H1 are significantly fainter, potentially due to severe synchrotron losses in a strong shock~\citep{meisenheimer86}. At high angular resolutions at centimeter wavelengths, the jet ridge oscillates around its mean position by up to 0.5\arcsec~\citep[e.g.,][]{conway93}.

In addition to its prominent total intensity structure, the polarization of the kiloparsec-scale jet has been extensively studied at radio and optical wavelengths~\citep[e.g.,][]{schmidt78,conway93,roeser96, perley17}. In \cite{hovatta19}, hereafter \citetalias{hovatta19}, we detected an extremely high Faraday rotation of $(5\pm0.3)\times10^5$\,\rad{} in the nucleus of 3C~273 over the 223 and 243\,GHz frequency range within ALMA band 6. Our modeling suggested that this is either due to multiple polarized components (each of which undergoes its own Faraday rotation) within the nucleus or due to internal Faraday rotation (when the jet Faraday-rotates its own polarized emission). Due to the limited wavelength coverage of our observations, we were unable to discriminate between the two scenarios. Additionally, because the ALMA polarization calibration could be performed reliably only for the inner third of the primary beam of the telescopes, we were unable to study the polarization structure of the kiloparsec-scale jet.

Here we present results from our ALMA full-polarization observations that were taken in 2018 using three ALMA bands (4, 6, and 7, around 145, 230, and 345\,GHz, respectively). We targeted both the nucleus and the kiloparsec-scale jet. Section~\ref{section:data} describes the ALMA observations and data reduction steps. Section~\ref{Section:results} gives the results from the analysis of the linear polarization in the core (Sect.~\ref{section:core}) and the kiloparsec-scale jet (Sect.~\ref{section:jet}).\ This is followed by a discussion in Sect.~\ref{section:discussion} and conclusions in Sect.~\ref{conclusions}.

\section{ALMA observations and data reduction}\label{section:data}
The Cycle 6 ALMA observations (project 2018.1.00357.S) were taken in bands 4, 6, and 7 in November 2018 using the Main Array (12-m antennas). Our observations used the full-polarization mode and the recommended setup for continuum polarization measurements. In each band, we used four 1.875\,GHz--wide spectral windows (Spws), each consisting of 64 channels that are 31.25\,MHz wide. This resulted in a total bandwidth of 7.5\,GHz spread over a much wider frequency range, 145--345 GHz.  Table~\ref{Table:obs} gives the band, date, duration, and Spw information for each observation. 

Because we were interested in the polarization of both the nucleus and kiloparsec-scale jet, we performed two separate pointings in each band: one centered on the nucleus (3C~273) and another on the bright hotspot H2 (3C273\_H2) about 21.3\arcsec{} away. In all bands, 3C~279 served as the bandpass and polarization calibrator, J1224+0413 as the check source, and J1224+0330 as the phase calibrator. However, because our target is much brighter than the phase calibrator, we used 3C~273 as the phase calibrator for itself (self-calibration, or self-cal) and for 3C273\_H2.

\begin{table*}
\centering
\caption{Observational setup. }
\label{Table:obs}
\begin{tabular}{l c c c c}
\hline
\hline
Band & Date & UT range & Spectral windows (GHz) & Beam size (\arcsec)\\
\hline
4 & 2018 Nov 18 & 12:40:09$-$16:26:47 & 138, 140, 150, 152 & $0.64\times0.51$ \\
6 & 2018 Nov 11 & 13:41:42$-$17:36:20 & 224, 226, 240, 242 & $0.38\times0.31$ \\
7* & 2018 Nov 15 & 12:36:36$-$14:38:08 & 336, 338, 348, 350 & $0.25\times0.22$ \\
7* & 2018 Nov 23 & 11:35:24$-$13:56:26 & 336, 338, 348, 350 & $0.23\times0.22$\\
\end{tabular}
\tablefoot{* In the analysis of the kiloparsec-scale jet the band 7 observations are concatenated together to improve the signal-to-noise ratio, and the beam size is $0.26\times0.22\arcsec$.}
\end{table*}

We performed the data reduction and imaging using Common Astronomical Software Applications (CASA) version 5.4.0. During calibration, we mostly followed the standard data reduction and polarization calibration steps \citep{nagai16}, with some exceptions in bands~6 and~7, as we describe below. In band~4, we set the amplitude scale of the observations using 3C~279 and assuming a flux density of 10.1\,Jy at 145\,GHz and spectral index of $-0.56$. We performed the calibration of the cross-hand delay, cross-hand phase, and instrumental polarization (D-terms) in the standard way, using 3C~279. 

In band 6, we also set the amplitude scale of the observations using 3C~279 and assuming a flux density of 8.47\,Jy at 230\,GHz and a spectral index of $-0.56$. The standard polarization calibration resulted in the imaginary part of the D-terms of the X and Y polarization
to have a symmetric slope with respect to zero as a function of frequency in Spws~2 and 3, which is unrealistic. Additionally, there were nonphysical jumps in the electric vector position angle (EVPA) values between the adjacent Spws. These effects are related to the use of a suboptimal model of the polarization calibrator, which was used to estimate both the instrumental polarization and the bandpass. To address this, we made two modifications to the standard calibration procedure. Firstly, we decided to adopt Event Horizon Telescope Collaboration's approach of setting the polarization model of the calibrator independently for each Spw, instead of using a single polarization model for all Spws~\citep[][]{goddi19}. This can mitigate the situations where the polarization calibrator has a significant rotation measure (RM) over the full bandwidth, reducing the spectral slope in the D-terms. Indeed, this approach significantly reduced --- though did not completely eliminate --- the D-term slope magnitude. We absorbed the remaining effect into our error estimates (see below).

Secondly, similar to~\citetalias{hovatta19}, we applied an additional self-bandpass calibration to the data after the polarization calibration steps. We first obtained a model of the source using CASA task \tclean, from which we found the total intensity of the bright point source in the nucleus to be 4.35\,Jy with a spectral index of $-0.85$, Stokes Q of 0.0238\,Jy, and Stokes U of $-0.1212$\,Jy; for this, we assumed that Stokes V vanishes (or is negligible) at a reference frequency of 232.86\,GHz. We then ran the bandpass estimation using 3C~273 as the calibrator, combining each scan and observation, normalizing the average solution amplitudes to 1.0, and applying the polarization calibration tables on the fly. This step ensures a correct amplitude ratio between polarizations at each frequency channel, resulting in EVPAs without abrupt jumps between adjacent Spws.

In band 7, we obtained two executions of the data because the first one did not reach the nominal 3-hour duration for full polarization tracks. In the first execution on 15 November 2018, we set the amplitude scale of the observations using 3C~279 assuming a flux density of 6.4\,Jy and spectral index of $-0.54$ at 343.5\,GHz, while on 23 November 2018 we assumed a flux density of 6.25\,Jy with a spectral index of $-0.56$. For this first execution, the real part of the D-terms showed two separate populations depending on the antennas, which is also unrealistic. We suspect that this was due to the relatively short duration of the execution (i.e., a limited parallactic-angle coverage for the polarization calibrator), resulting in a degraded quality of the D-term estimates. Given the short time between executions at the same band, we assumed similar D-terms for the two datasets, and copied the D-term solutions from the second execution (where the D-terms did not show any pathological behavior) into the first. There are, however, five antennas that were present on 15 November and were not available on 23 November. For those antennas, we did not correct for instrumental polarization. After producing the images of the two executions separately, there were no changes in the EVPAs of the brightest hotspot, indicating a consistent calibration of the two epochs. However, a $10\deg$ shift in the EVPAs of the nucleus was found, showing evidence of an intrinsic EVPA variability of the nucleus.  Data of 3C~273 from the AMAPOLA\footnote{\url{https://www.alma.cl/~skameno/AMAPOLA/J1229+0203.flux.html}} grid survey of calibrator sources \citep{kameno23} show that the EVPA at band 7 is stable from 11$-$17 November while a clear jump, consistent with our data, is seen in the next grid survey observations on 28 November. Therefore, in the following analysis of the nucleus, we only considered the 15 November epoch, which is closer in time to the other bands, while for the analysis of the kiloparsec-scale jet  where no variability is seen, we combined  the epochs to improve the signal-to-noise ratio. 

\subsection{Extracting Stokes parameters of the nucleus}
Because the bright nucleus of 3C~273 dominates the emission, we applied self-cal to improve the dynamic range of our data. For this, we first obtained a model of the source using the CASA task \tclean, and then applied a round of phase-only self-cal with solution interval set to \emph{int}. We then improved the model and applied a round of phase and amplitude self-cal with the solution interval set to 10~\si{s} in bands 4 and 6 and 30~\si{s} in band 7. We repeated this procedure twice for each band. 

Similar to \citetalias{hovatta19}, we extracted the Stokes parameters of the nucleus using the external CASA library \texttt{UVMULTIFIT} \citep{marti-vidal14}. We assumed that a single point source dominates the emission at the phase center of the observation and fit the visibility data with a delta component centered on each 31.25\,MHz-wide Spw independently. Following \citetalias{hovatta19}, we ran Monte Carlo (MC) simulations to account for thermal and systematic uncertainties in the data, and obtained a distribution of 1000 Stokes I, Q, U, and V values. We describe the MC procedure in Appendix A of \citetalias{hovatta19} and give various noise contributions for each band in Table~\ref{Table:noise}. Note the higher noise contributions for the D-terms in band 6 and the 15 November execution of band 7 due to the issues described above.

We note that the resulting EVPA values from the MC simulations slightly differ from the values obtained using CASA images (up to $1\deg$ in band 6) due to a different way of calculating the parallactic angle in CASA versus\hbox{} our MC script. This is why we consistently use the values obtained from the MC simulations in our analysis. We show the total intensity (Stokes $I$), the normalized Stokes parameters, $q = Q/I$ and $u = U/I$, fractional polarization, $p=\sqrt{q^2+u^2}$, and the polarization angle, $\psi=0.5\arctan(U/Q)$  of the unresolved core as a function of wavelength-squared, $\lambda^2$, for all the bands in Fig.~\ref{fig:m4_specfit}. The emission over all the bands is optically thin  with a spectral index of $\alpha_{\rm core}=-0.7915\pm0.0003$ and a reduced $\chi^2=2.76$ of the fit when using the 15 November epoch in band 7. If we use the 23 November epoch instead, the spectral index would be $-0.905\pm0.002$ with a significantly higher reduced $\chi^2=123.0$, further supporting our decision to use only the 15 November epoch in the modeling of the core polarization. The polarization spectrum of the core is complex, and we return to the modeling of the polarization in Sect.~\ref{section:core}. We give the mean and standard deviation of the Stokes I, Q, and U distributions for each channel in an online table, available at the CDS. Column 1 gives the observing date, Column 2 the frequency of the channel, Columns 3$-$8 the Stokes I, Q, U values and their uncertainties, and Columns 9$-$12 the fractional polarization and EVPA with their uncertainties.

\begin{table*}
\centering
\caption{Parameters of the Gaussian noise contributions in our MC simulations.}
\label{Table:noise}
\begin{tabular}{l l l l l}
\hline
\hline
Noise Type & Band 4 & Band 6 & Band 7 Nov 15 & Band 7 Nov 23\\
\hline
X-Y Phase [deg.] & 2 & 2  & 2 & 2 \\
X-Y Bandpass amplitude [\%] & 0.1 & 0.1 & 0.2 & 0.2 \\
X-Y Bandpass phase [deg.] & 0.5 & 0.5 & 0.5 & 0.5 \\
Thermal [mJy] & 0.16 & 0.44 & 0.14 & 0.075 \\
D-terms (real \& imag) [\%] & 0.5 & 2 & 1 & 0.5 \\
\end{tabular}
\tablefoot{We assume a mean value of $0$ for each parameter and report  the standard deviation in this table. Here, ``X-Y Phase'' is the residual phase between the polarizers at the reference antenna, ``X-Y bandpass'' is the (channel-wise) residual noise of the bandpass, relative between polarizers, ``Thermal'' is the theoretical image sensitivity in the continuum (as estimated from the ALMA Sensitivity Calculator), and ``D-terms'' is the (channel-wise) residual polarization leakage.}
\end{table*}

\subsection{Imaging and alignment of the kiloparsec-scale jet}
We obtained the images of the kiloparsec-scale region in each band using the CASA task \tclean. The bright nucleus of 3C~273 is present in bands~4 and~6, providing a strong point source suitable for self-calibration, which we can use to improve the dynamic range of the resulting images. For this, we followed an approach similar to the previous section. However, in band 7, the image does not show the nucleus due to the smaller primary beam, preventing us from using self-cal. In bands 4 and 6, we obtained the final aggregated bandwidth images of all Spws by using specmode \emph{mfs}, deconvolver \emph{mtmfs}, and setting \texttt{nterms} to 2 to account for the wide spectral coverage. In band~7,  due to the smaller fractional bandwidth, we obtained the aggregated bandwidth images using specmode \emph{mfs} and deconvolver \emph{hogbom}.  Additionally, in all bands, we imaged each Spw independently. We also obtained image cubes for each Spw, but the sensitivity in each 31.25\,MHz channel in the hotspot region was not high enough to use these for further analysis. We used the task \emph{immath} to extract the Stokes I, Q, and U images for further processing and plotting. Table~\ref{Table:obs} gives the resulting beam size (corresponding to the angular resolution) for each of the bands.

In order to combine the images across the three bands, we also produced images of each Spw that we smoothed using the \emph{imsmooth} task to have the same beam size (which corresponds to the beam size of the Spw 0 in band 4, $0.54\arcsec\times0.44\arcsec$). Because the self-cal we applied changed the absolute position of the phase center in our images, we used the quality assurance level 2 (QA2) images delivered within the dataset (with no self-cal applied) to verify the image alignment between different bands. This resulted in a small shift of $-1$ pixel ($-0.043\arcsec$) in the declination of the band 6 kiloparsec-scale images. We then used these smoothed images to produce the spectral index maps in Sect.~\ref{Section:results}. 

\section{Results}\label{Section:results}
In this section we present the polarization and Faraday-rotation properties of both the compact core and the kiloparsec-scale jet of 3C~273, derived from our ALMA band~4, 6, and 7 observations. The broad wavelength coverage (145--345 GHz) allows us to revisit the extreme Faraday rotation previously detected in band~6 alone \citepalias{hovatta19} and to better test whether the behavior arises from multiple polarized components or from internal Faraday rotation within the nuclear region. For the extended kiloparsec-scale jet, we analyzed the resolved linear polarization in both transverse and longitudinal profiles, and not the Faraday rotation, which produces negligible EVPA rotations at these frequencies.

\subsection{Polarization in the core}\label{section:core}
To characterize the frequency-dependent behavior of the unresolved core, we modeled the complex fractional polarization as a function of wavelength-squared:
\begin{equation}\label{eq:pl2_def}
p(\lambda^2) \equiv q(\lambda^2) + i\,u(\lambda^2),
\end{equation}
where $q$ and $u$ are the fractional Stokes parameters. 
We performed this QU-fitting in the complex plane using parametric models,
$p(\lambda^2;\,\bm{\theta})$, and a nested-sampling backend 
that returns posterior samples for the parameter set $\bm{\theta}$ 
and the Bayesian evidence $\ln\mathcal{Z}$ for model comparison.

\begin{figure*}
\centering
\includegraphics[width=\textwidth]{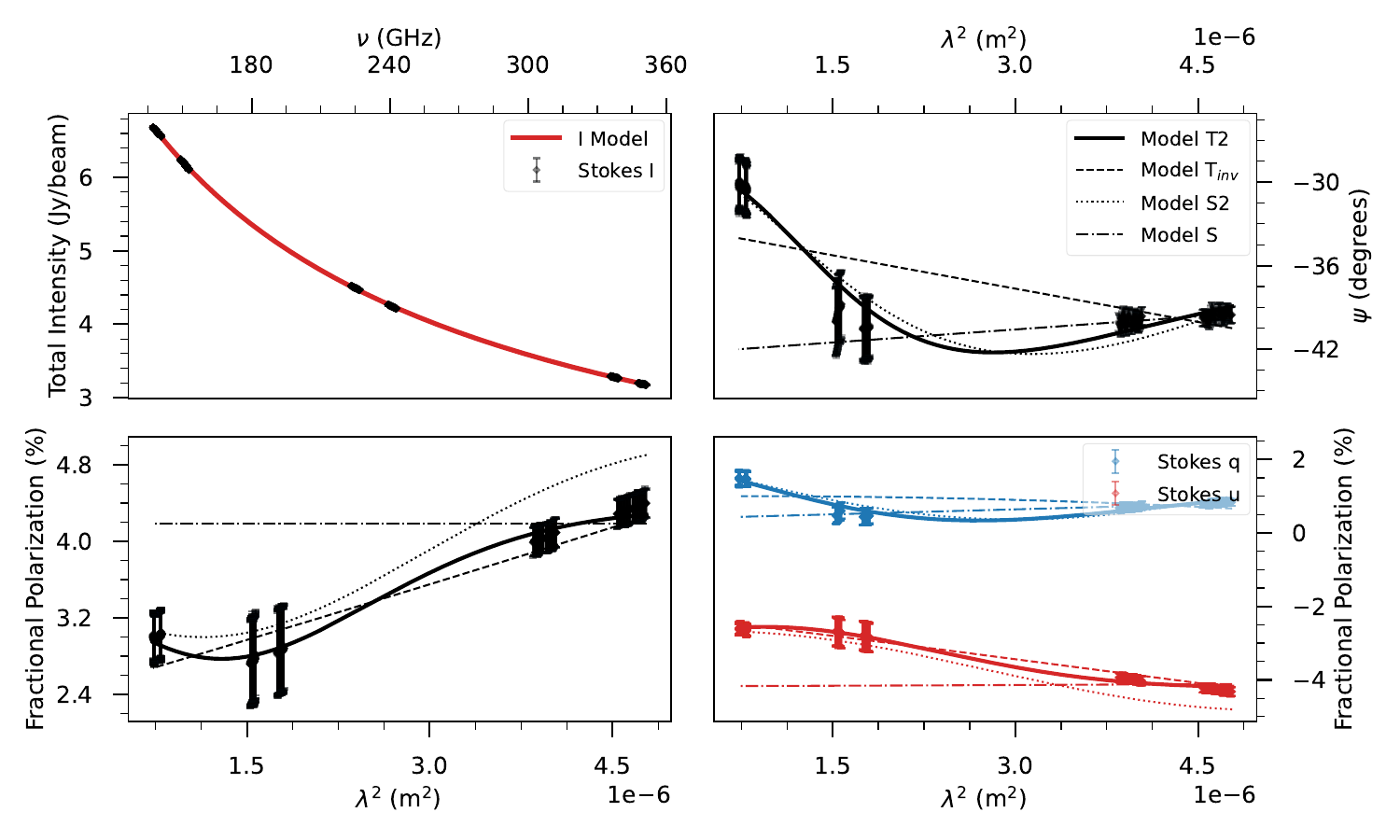}
\caption{Data and model curves for the four models tested across ALMA Bands~4, 6, and 7  of the unresolved core.
\textit{Top left}: Stokes $I$ vs.\ frequency. \textit{Top right}: Polarization angle ($\psi$) vs.\ $\lambda^{2}$.
\textit{Bottom left}: Fractional polarization ($p$) vs.\ $\lambda^{2}$. \textit{Bottom right}: Stokes $q$ and $u$ vs.\ $\lambda^{2}$. The thick solid lines correspond to the best-fit model, which is the two-component external-dispersion model (T2). The dashed curve is the inverse depolarization (T$_\mathrm{inv}$), the dotted curve is the thin two-Faraday components (S2), and the dot-dashed curve is a single Faraday-thin component (S).
The T2 model captures both the frequency-dependent depolarization and the smooth EVPA rotation across the bands.
}
\label{fig:m4_specfit}
\end{figure*}

We investigated four distinct models for the core polarization data 
from bands 4, 6, and 7 (15 November 2018 epoch). We started with the simplest model~S, which corresponds to a single 
Faraday-thin component. We also considered three other models that can in principle describe an increasing fractional polarization with wavelength (i.e.,~inverse depolarization): (i)~model~T\textsubscript{inv}, which represents a single inverse-depolarizing Faraday component; (ii)~model~S2, which represents two Faraday-thin components; and (iii)~model~T2, which represents two external dispersive Faraday components.
We summarize these models
in Table~\ref{tab:model_summary} and give their functional forms in Appendix~\ref{app:model_defs}. The data and the model fits are shown in Fig.~\ref{fig:m4_specfit}. 

\begin{table}
\centering
\caption{Polarization models.}
\label{tab:model_summary}
\begin{tabular}{l c c}
\hline
\hline
Model (alias) & \# parameters & Eq. \\
\midrule
S (single thin)             & 3  & \ref{eq:m1_model} \\
T\textsubscript{inv} (single inverse-depol) 
                            & 5  & \ref{eq:m5_model} \\
S2 (double thin)            & 6  & \ref{eq:m11_model} \\
T2 (double ext.\ dispersion)& 8  & \ref{eq:m4_model} \\

\end{tabular}
\end{table}

We compared the models via their Bayesian evidence, $\ln\mathcal{Z}$, defining the Bayes factor as
\begin{equation}\label{eq:bayes_factor}
\mathcal{B}_{a,b} = \ln\!\left(\frac{\mathcal{Z}_a}{\mathcal{Z}_b}\right)
= \ln(\mathcal{Z}_a) - \ln(\mathcal{Z}_b),
\end{equation}
where $\mathcal{B}_{a,b}\gtrsim5$ indicates a very strong preference for model $a$ 
\citep{Kass1995}. Table~\ref{tab:qufit_ranking} gives the Bayesian evidence and relative ranking for each model.

\begin{table}
\centering
\caption{Model ranking ordered by $\ln\mathcal{Z}$ (highest first). }
\label{tab:qufit_ranking}
\begin{tabular}{lrrrl}
\hline
\hline
Model & $\ln\mathcal{Z}$ & $\Delta\ln\mathcal{Z}$  & Ruled out & $\chi^2_{\rm r}$\\
\midrule
T2  & 6437 & 0.00   & --- (best)  & 0.35  \\
S2 & 6296 & 141 & decisive & 0.6  \\
T\textsubscript{inv}  & 5282 & 1155& decisive     & 2.1  \\
S  & 2134 & 4302 & decisive    & 23.2 \\
\end{tabular}
\tablefoot{The Bayes factor is $\Delta\ln\mathcal{Z}=\ln\mathcal{Z}_{\rm best}-\ln\mathcal{Z}_{\rm model}$. The ``Ruled out'' column rates the evidence for the best model over each alternative (i.e.,~all models are decisively ruled out compared to the  best fit model).}
\end{table}

Table~\ref{tab:qufit_ranking} shows that among the four tested models, the two–component external–dispersion case (T2) provides 
the highest Bayesian evidence and the lowest reduced $\chi^2$. 
This indicates a good representation of both the polarization fraction, $p(\lambda^2)$, and EVPA, $\psi(\lambda^2)$, across ALMA bands~4, 6, and 7.  Table~\ref{table:T2param} gives the best-fit parameters for this model while the parameters for the other models are given in Appendix~\ref{app:params}. 
The simpler two–thin component model (S2) gives an inferior but acceptable fit based on the reduced $\chi^2$ but is decisively ruled out based on the Bayes Factor ($\Delta\ln\mathcal{Z}\simeq140$). As can be seen in Fig.~\ref{fig:m4_specfit}, it fails to reproduce the fractional polarization behavior. The single–component models (T\textsubscript{inv}, S) cannot reproduce the observed fractional polarization or EVPA behavior and are also decisively disfavored. 

\begin{table}
\centering
\caption{Best-fitting parameters for the two-component external Faraday-dispersion (T2) model, with reduced $\chi^2$ and Bayesian evidence.}
\label{table:T2param}
\begin{tabular}{lc}
\hline\hline
Parameter & Value \\
\midrule
$p_1$ (\%) & $2.03^{+0.05}_{-0.06}$ \\
$p_2$ (\%) & $4.31^{+0.02}_{-0.02}$ \\
$\psi_{0,1}$ (deg) & $20.5^{+1.1}_{-1.1}$ \\
$\psi_{0,2}$ (deg) & $138.05^{+0.59}_{-0.57}$ \\
RM$_1$ (rad m$^{-2}$) & $(2.58^{+0.08}_{-0.08})\times 10^{5}$ \\
RM$_2$ (rad m$^{-2}$) & $(1.48^{+0.18}_{-0.18})\times 10^{4}$ \\
$\sigma_{\mathrm{RM},1}$ (rad m$^{-2}$) & $(2.36^{+0.07}_{-0.07})\times 10^{5}$ \\
$\sigma_{\mathrm{RM},2}$ (rad m$^{-2}$) & $(2.30^{+2.52}_{-1.63})\times 10^{3}$ \\
\midrule
$\chi^2_r$ & 0.35 \\
$\ln Z$ & $6437$ \\
\hline
\end{tabular}
\end{table}

The individual parameter values for the best-fitting T2 solution are shown in Fig.~\ref{fig:m4_corner}. It consists of two polarized components: one with a high RM of RM$_1 = +(2.58\pm0.08)\times10^5$\rad~and comparably large Faraday dispersion of $\sigma_{\rm RM_1}=(2.36\pm0.07)\times10^5$\rad, and the other with a moderate RM$_2 = +(1.5\pm0.2)\times10^4$\rad~and a 2-sigma upper limit on the dispersion of $\sigma_{\rm RM_2}<0.7\times10^4$\rad. 
The associated intrinsic degree of polarizations of $p_{0,1}\sim2\%$ and $p_{0,2}\sim4.3\%$ are of the total flux, so components 1 and 2 correspond to intrinsic polarized fluxes of $\sim$80~mJy and $\sim$170~mJy, respectively. 
The intrinsic polarization angle of component 2 is approximately perpendicular to the parsec-scale jet direction of $\sim-130^\circ$ \citep{hovatta12} indicating a magnetic field orientation parallel to the jet, which is consistent with the field orientation in inner part of the jet observed in \citep{asada02}. The intrinsic angle of component 1 is at an oblique angle of $\sim-30^\circ$ to the parsec-scale jet direction.

While the best-fit-model equation is nominally of external Faraday rotation, we note that in the case of unresolved emission we cannot reliably distinguish this from internal Faraday rotation effects that would cause similar spectro-polarimetric behavior. Therefore, we consider this model as a parameterization of mean RM and dispersion, and the intrinsic polarization properties. We also note that as the data in the different bands are not strictly simultaneous, and we see significant variability in the band 7 EVPA on a timescale of a week, there is additional uncertainty in the best-fit model due to potential variability. Therefore, we do not consider it useful to fit more complicated models (or equally complicated models with different underlying assumptions) in the absence of more robust prior information based on contemporaneous very long baseline interferometry (VLBI) imaging.

\subsection{Polarization and spectral index of the kiloparsec-scale jet}\label{section:jet}
\begin{figure}
    \centering
    \includegraphics[trim=0.8cm 2cm 1cm 2cm,clip=true,width=1.0\linewidth]{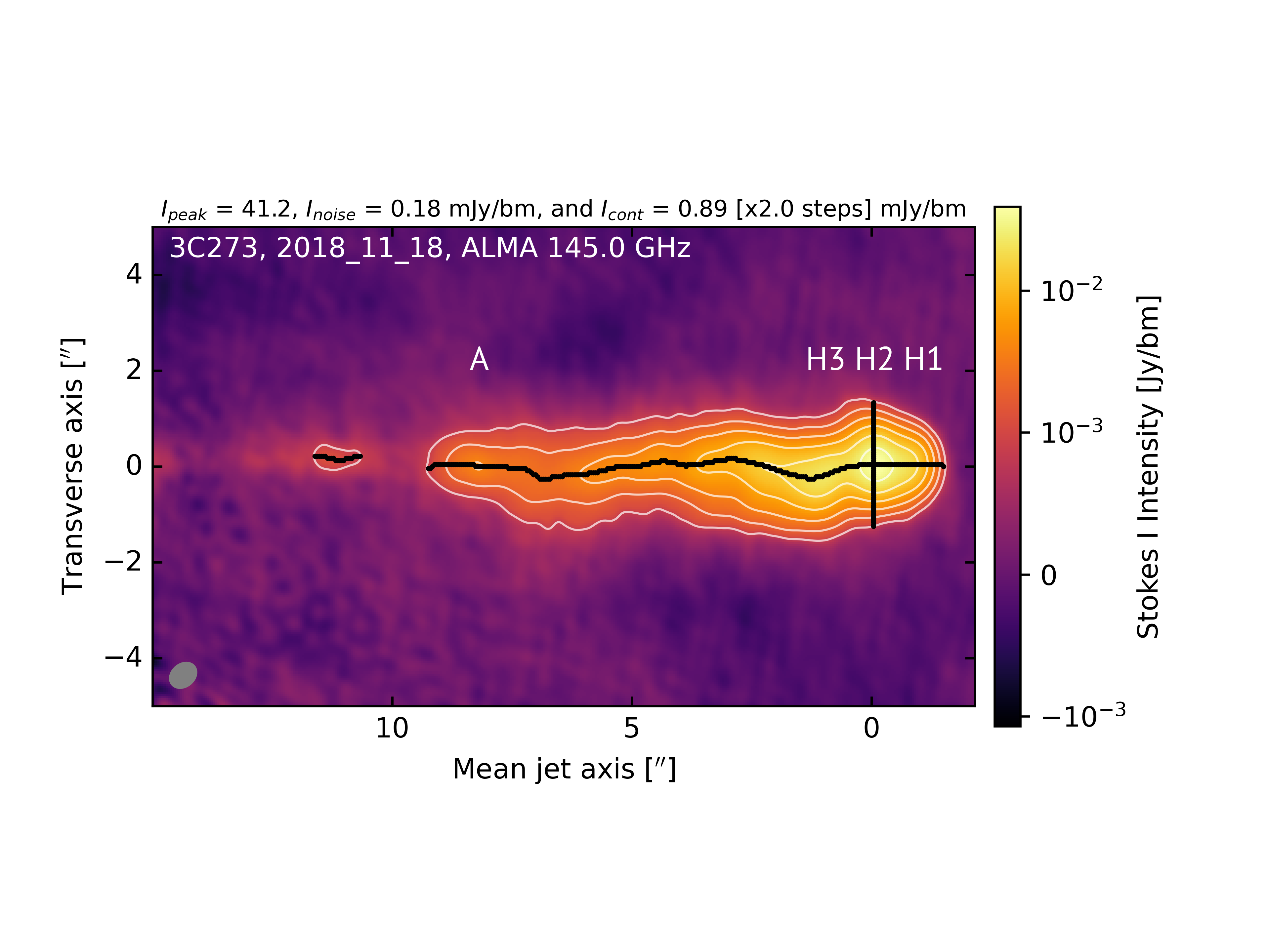}
    \begin{minipage}{\linewidth}
    \raggedright
    \includegraphics[width=0.89\linewidth]{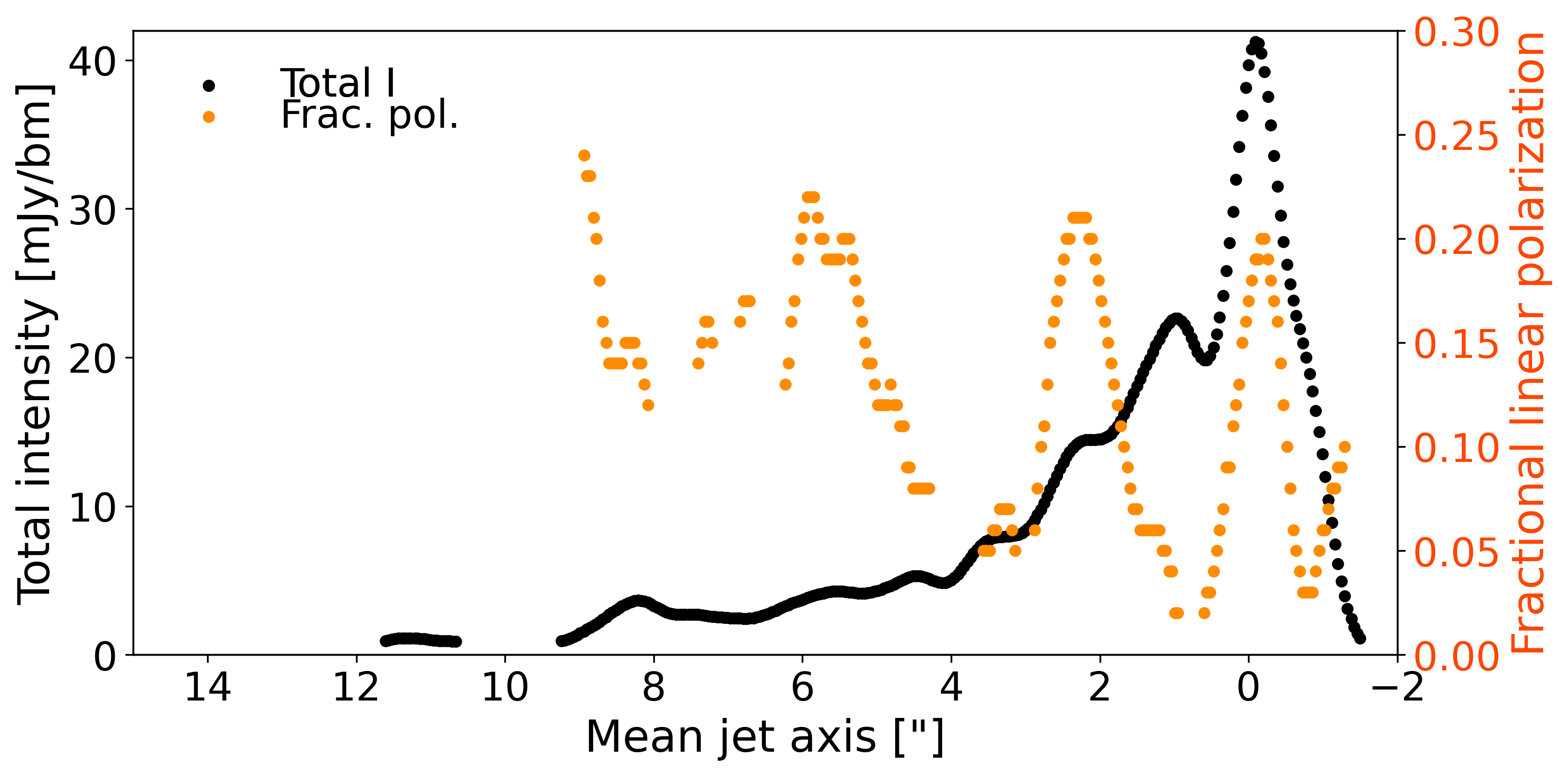}
    \end{minipage}
    \includegraphics[trim=0.8cm 2cm 1cm 2cm,clip=true,width=1.0\linewidth]{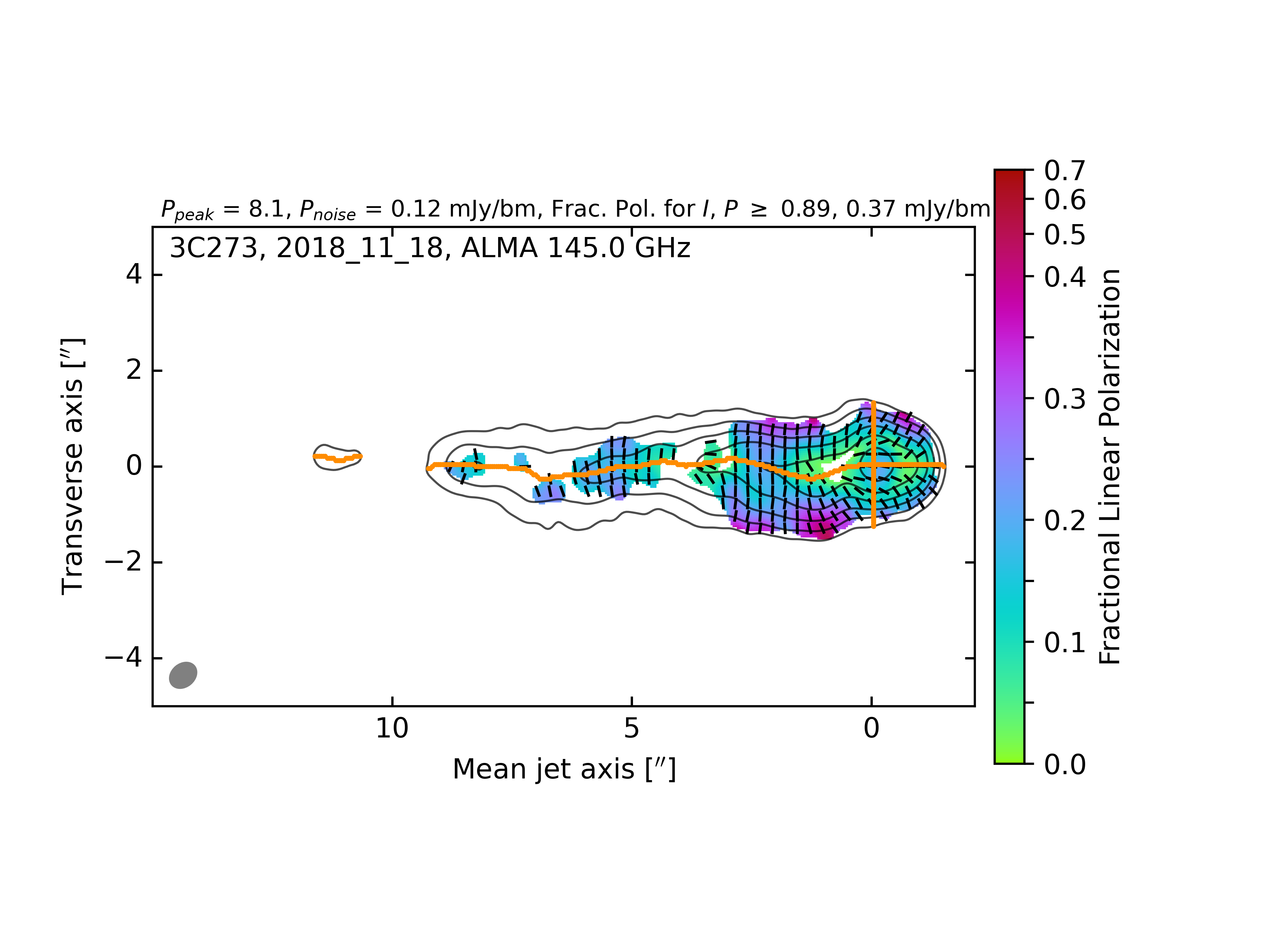}
    \caption{Total intensity (\textit{top}) and fractional linear polarization (\textit{bottom}) of the outer kiloparsec-scale jet in ALMA band 4.  The locations of the most prominent knots are marked in the total intensity map. In both panels the contours indicate the total intensity starting at  five times the image rms level (given above the maps) and increasing in steps of 2.0. The total intensity ridge line along the jet and a slice across the brightest pixel in the hotspot region is plotted in black (top) and orange (bottom). \textit{Middle}: Total intensity (black, left y-axis) and fractional polarization (orange, right y-axis) values along the jet ridge line.}
    \label{fig:hotspotB4}
\end{figure}

\begin{figure}
    \centering
    \includegraphics[trim=0.8cm 3cm 1cm 3cm,clip=true,width=1.0\linewidth]{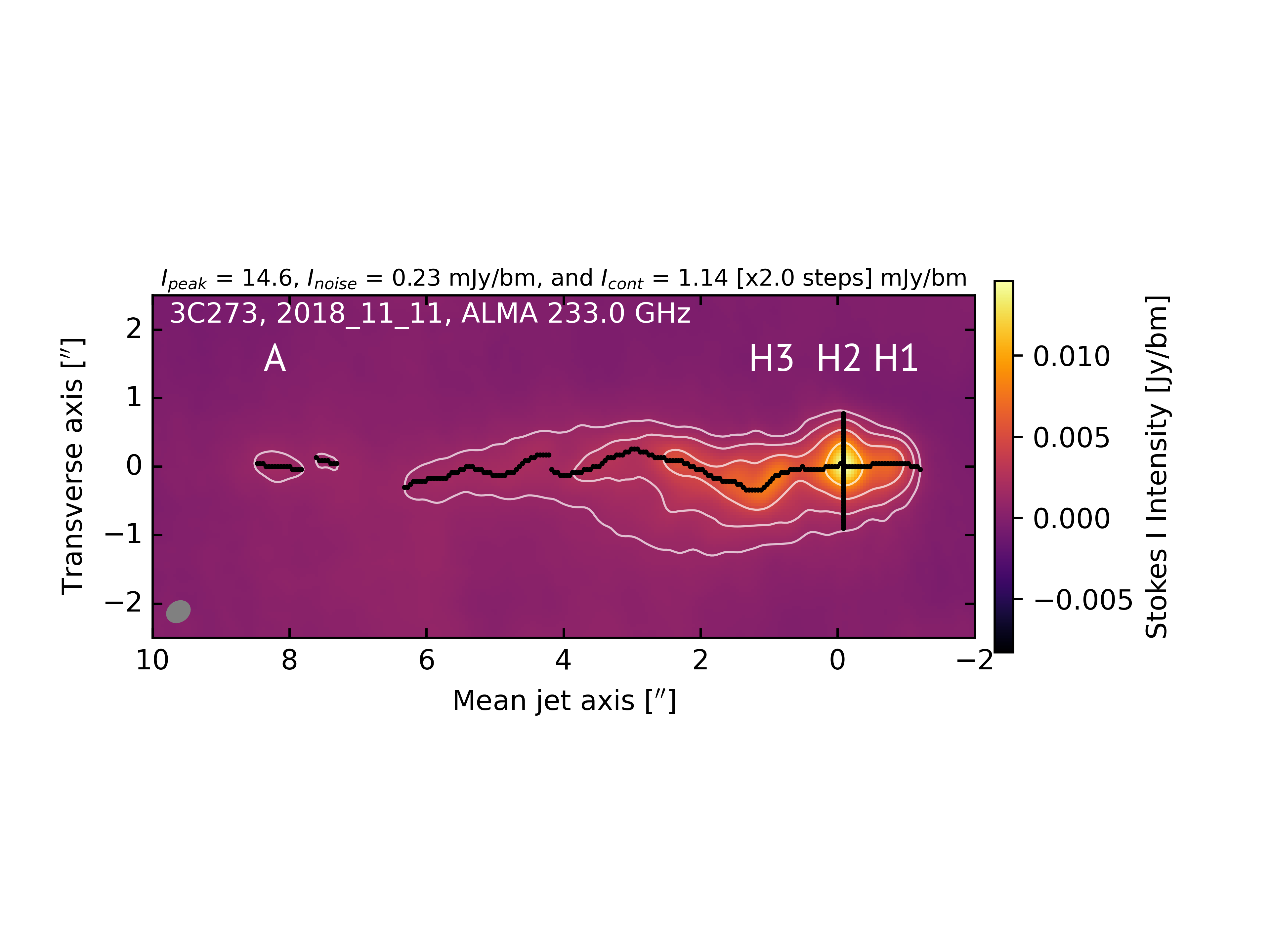}
    \begin{minipage}{\linewidth}
    \raggedright
    \includegraphics[width=0.89\linewidth]{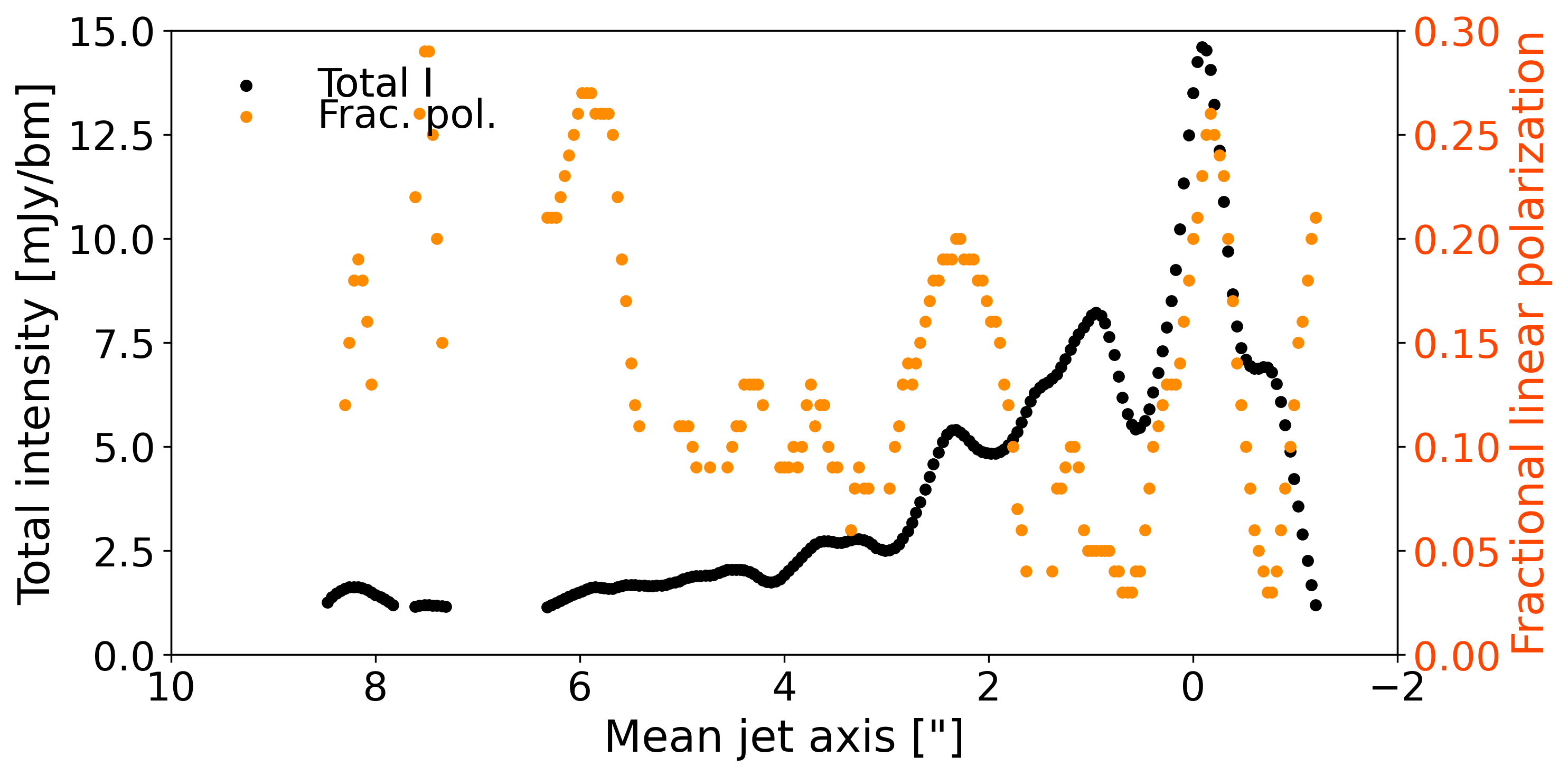}
    \end{minipage}
    \includegraphics[trim=0.8cm 3cm 1cm 3cm,clip=true,width=1.0\linewidth]{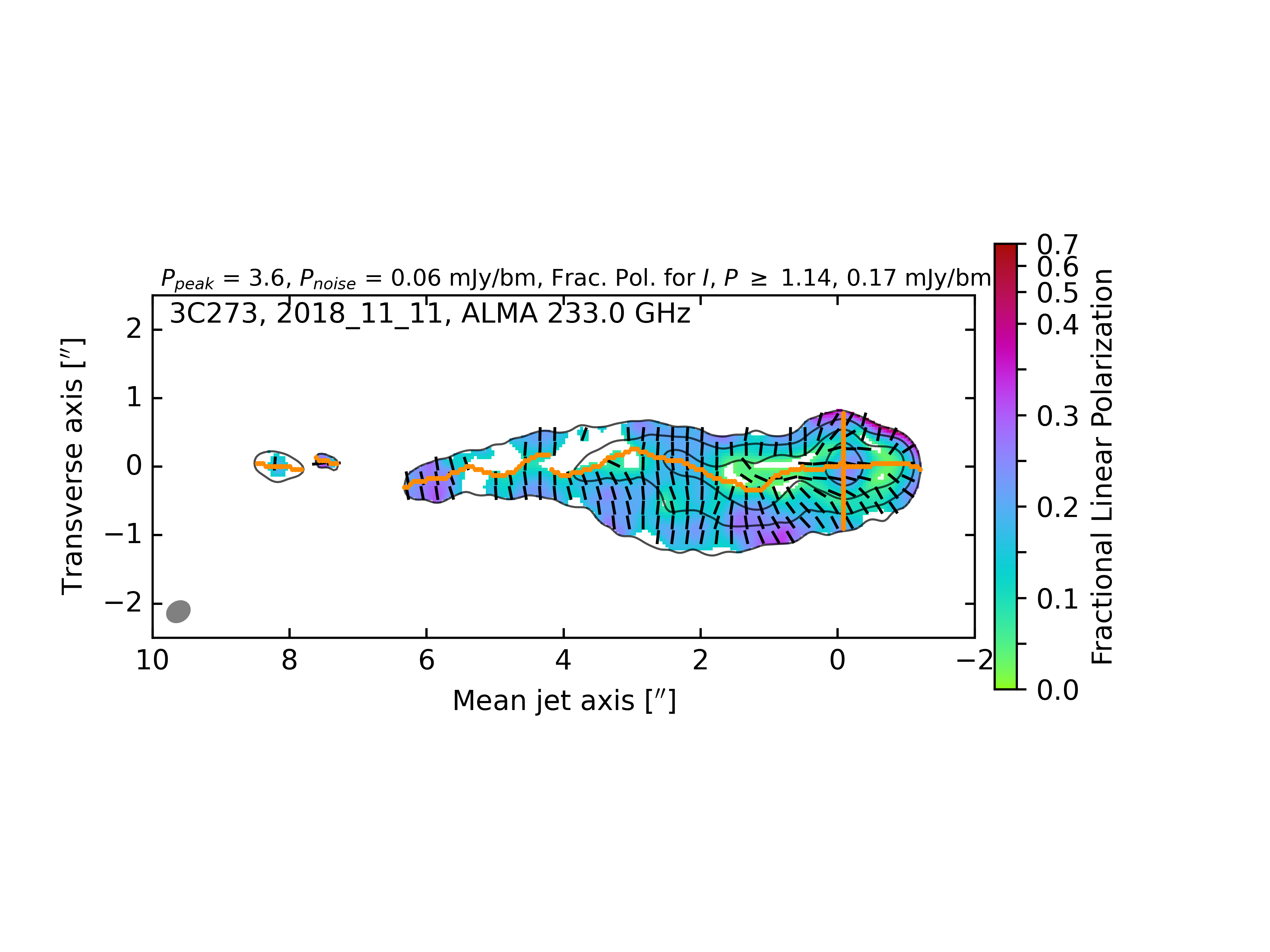}
    \caption{Same as Fig.~\ref{fig:hotspotB4} but for ALMA band 6.}
    \label{fig:hotspotB6}
\end{figure}

\begin{figure}
    \centering
    \includegraphics[trim=0.8cm 2cm 1cm 2cm,clip=true,width=1.0\linewidth]{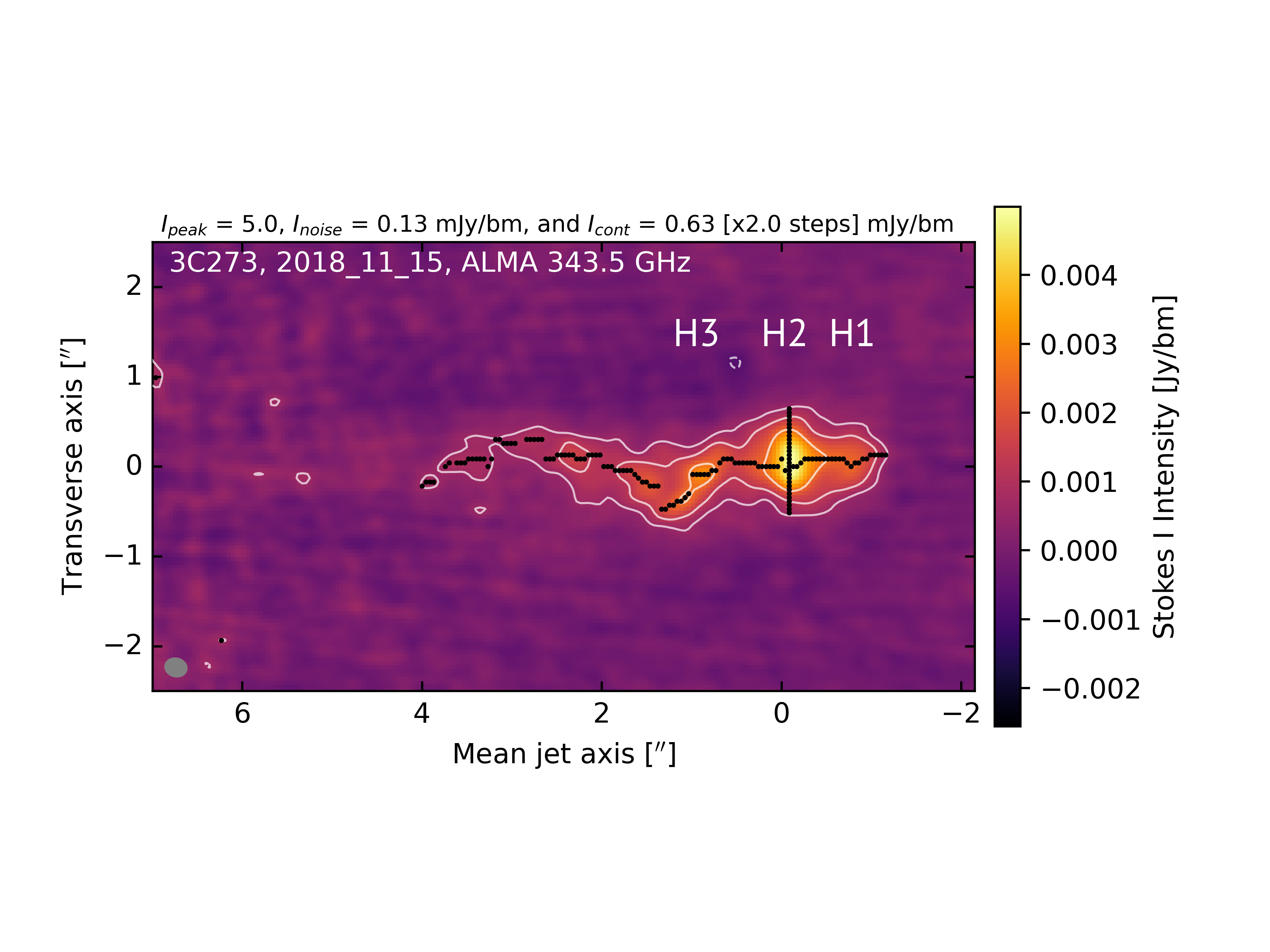}
    \begin{minipage}{\linewidth}
    \raggedright
    \includegraphics[width=0.90\linewidth]{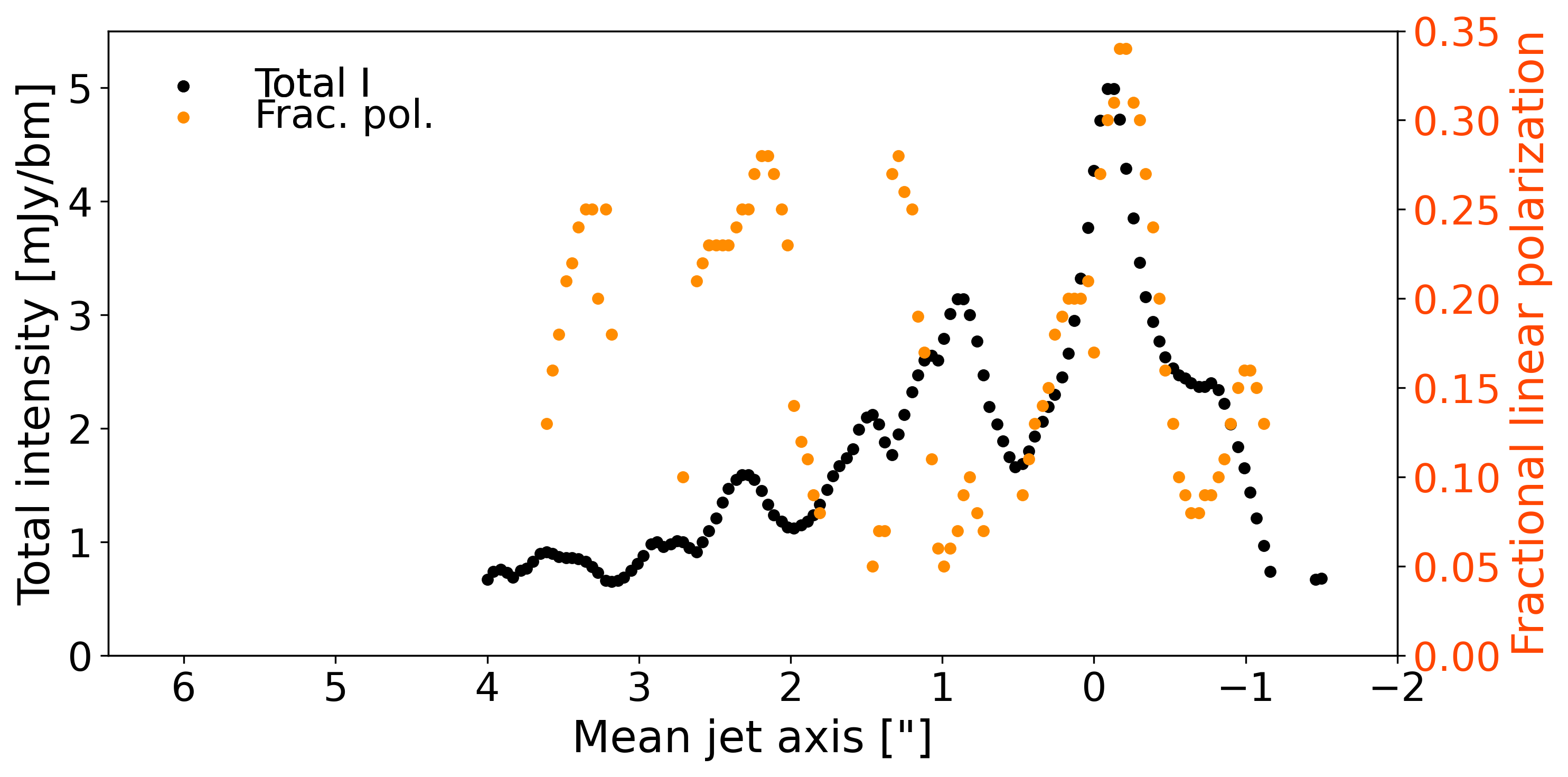}
    \end{minipage}    
    \includegraphics[trim=0.8cm 2cm 1cm 2cm,clip=true,width=1.0\linewidth]{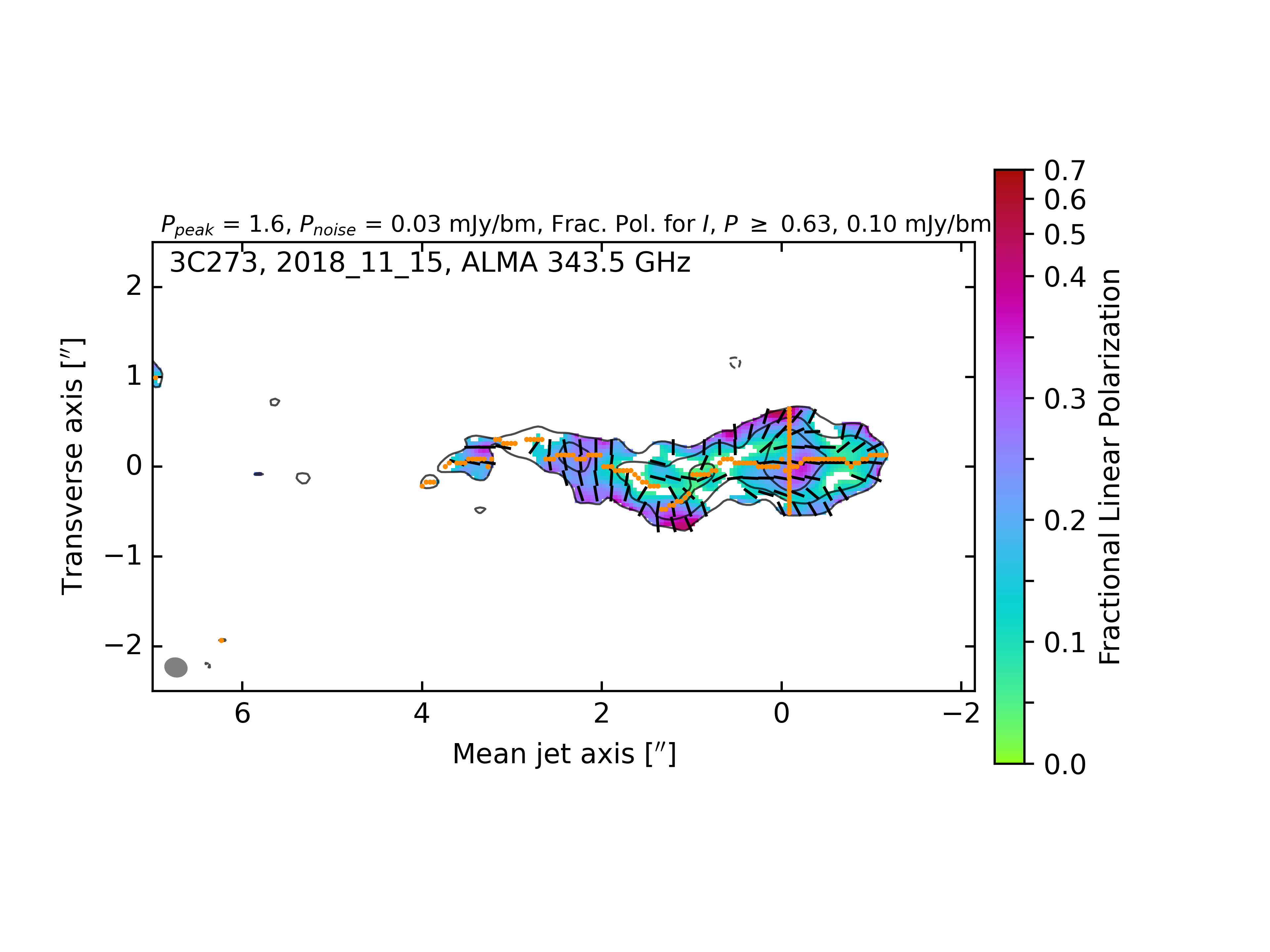}
    \caption{Same as Fig.~\ref{fig:hotspotB4} but for ALMA band 7. }
    \label{fig:hotspotB7}
\end{figure}

Images of the total intensity ($I$) and fractional polarization ($p$) of the kiloparsec-scale jet for each band are shown in Figs.~\ref{fig:hotspotB4}-\ref{fig:hotspotB7}. These are rotated by $48\deg$ for visualization purposes. 
In  all bands, the fractional polarization was calculated for all pixels where the total intensity exceeded five times its rms uncertainty and the polarized flux density exceeded three times its rms uncertainty.
The rms values are given in the figures.

\begin{figure}
    \centering
    \includegraphics[width=1.0\linewidth]{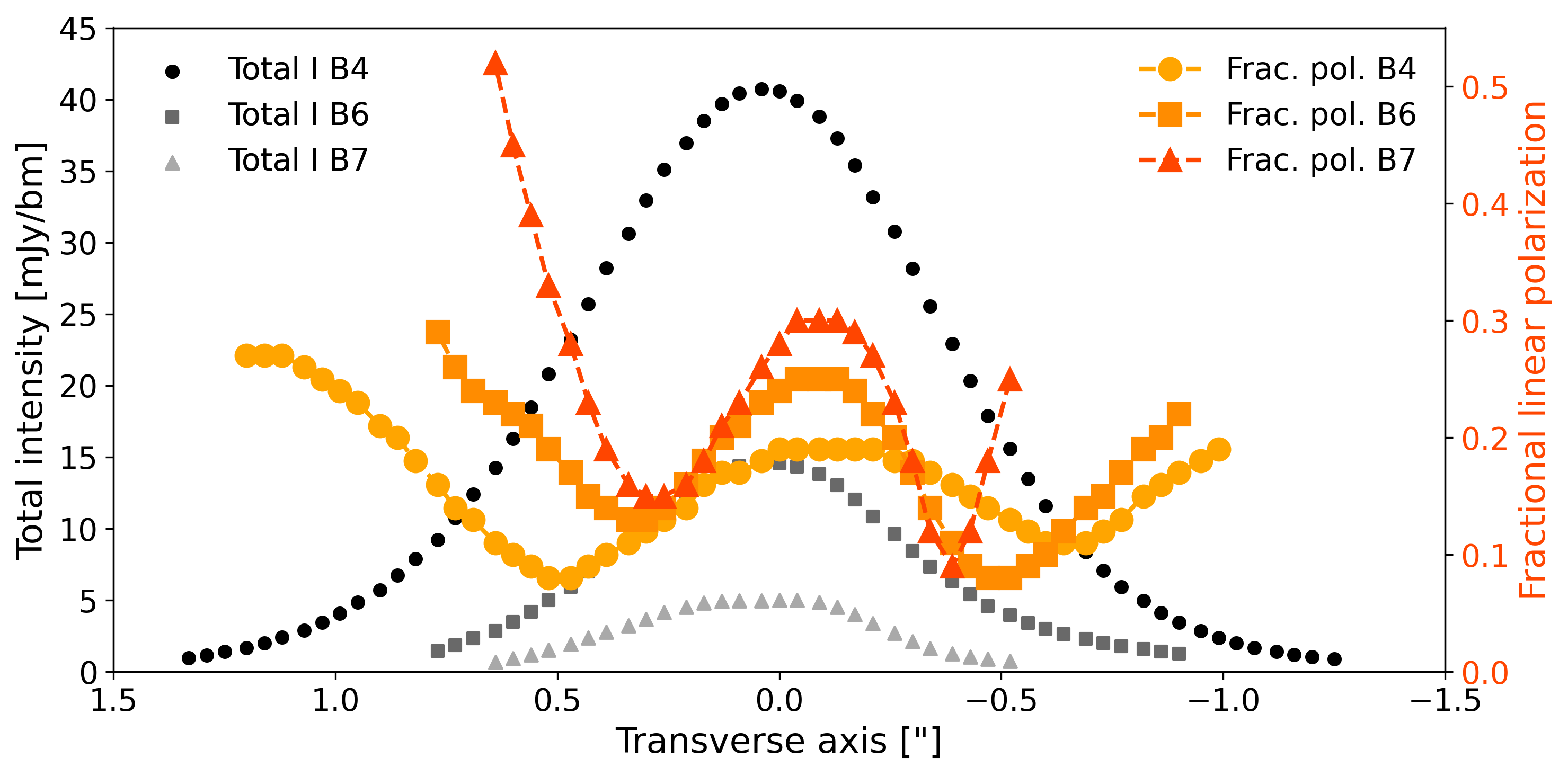}
    \caption{Total intensity (left y-axis) and fractional polarization (right y-axis) across the jet along the slice over the brightest hotspot pixel for each band.}
    \label{fig:slices}
\end{figure}

The directions of the EVPA vectors are also plotted on the fractional polarization images. We also obtained the total intensity ridge line of the jet by extracting the location of the brightest pixel along the jet in  the mean jet axis direction. Similarly, we extracted a slice across the jet in  the transverse axis direction at the position of the brightest  hotspot pixel. We show the total intensity and fractional polarization values along the ridge lines in Figs.~\ref{fig:hotspotB4}-\ref{fig:hotspotB7} and along the slices in Fig.~\ref{fig:slices}.

As expected for optically thin synchrotron emission, the total intensity is the highest at 145\,GHz, where we detect all the main jet components, starting from knot A. At 233\,GHz knot A is barely detected while at 344\,GHz we only see the hotspot region starting about 4\arcsec inward from H2. The fractional polarization, on the other hand, is somewhat higher at the higher frequencies, but this could be due to their better angular resolution resulting in less beam depolarization. Both the total intensity and the fractional polarization oscillate along the ridge line, with multiple locations with a higher total intensity and fractional polarization. These are the most pronounced at 344\,GHz where also the angular resolution is the best and the ridge line follows the wiggles of the jet spine more closely.

In all the bands, the EVPA vectors at the hotspot location are parallel to the jet indicating that the magnetic field is perpendicular to the jet. Toward the edges of the jet, the direction changes by about 90 degrees so that at the jet edges the magnetic field is parallel to the jet. This is also reflected in the fractional polarization across the slices shown in Fig.~\ref{fig:slices}, where there are two minima at the locations where the EVPA changes direction and the orthogonal polarization components cancel each other. We come back to this point in Sect.~\ref{section:discussion}.

\begin{figure}
    \includegraphics[width=1.0\linewidth]{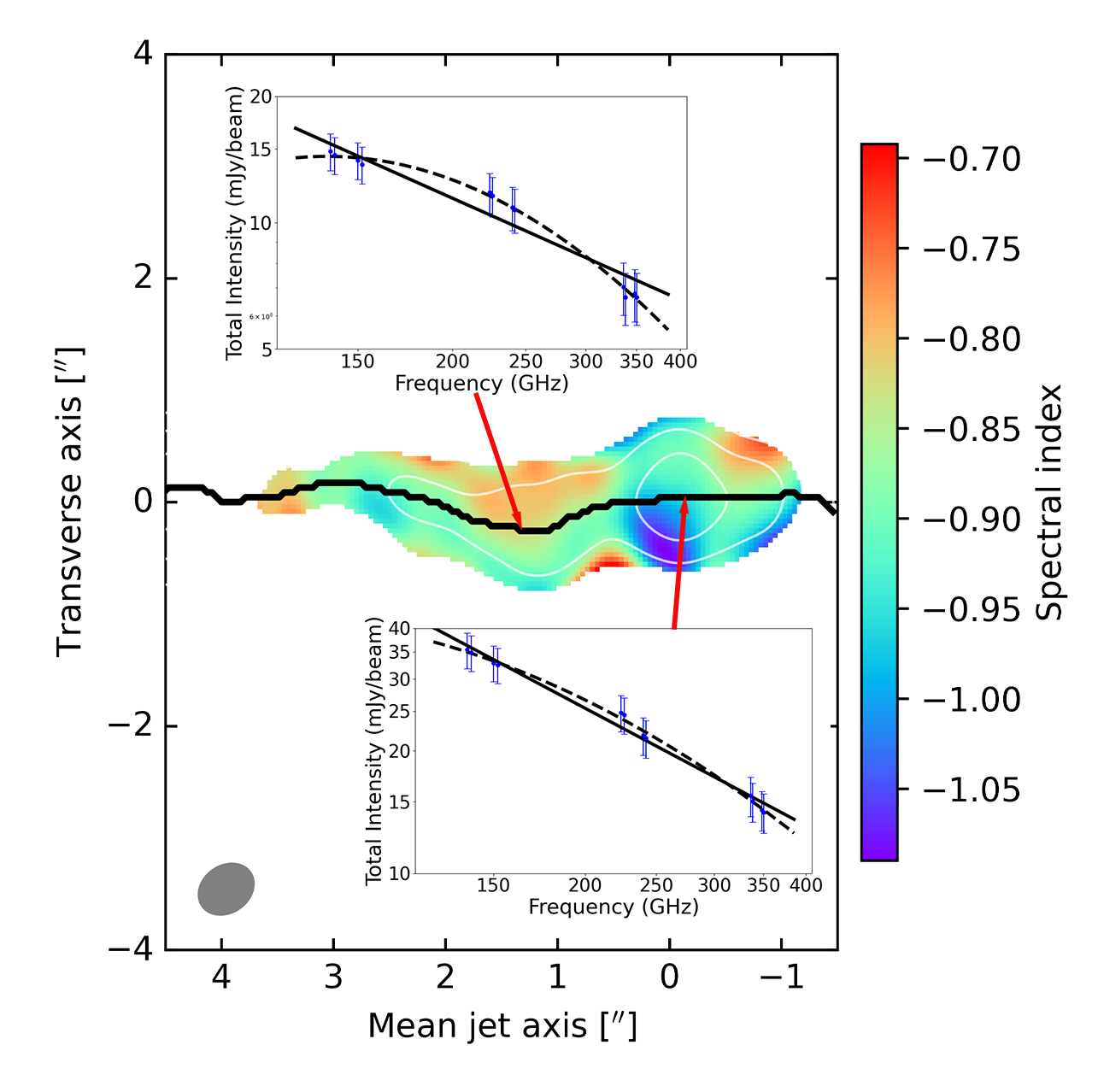}
    \includegraphics[width=0.8\linewidth]{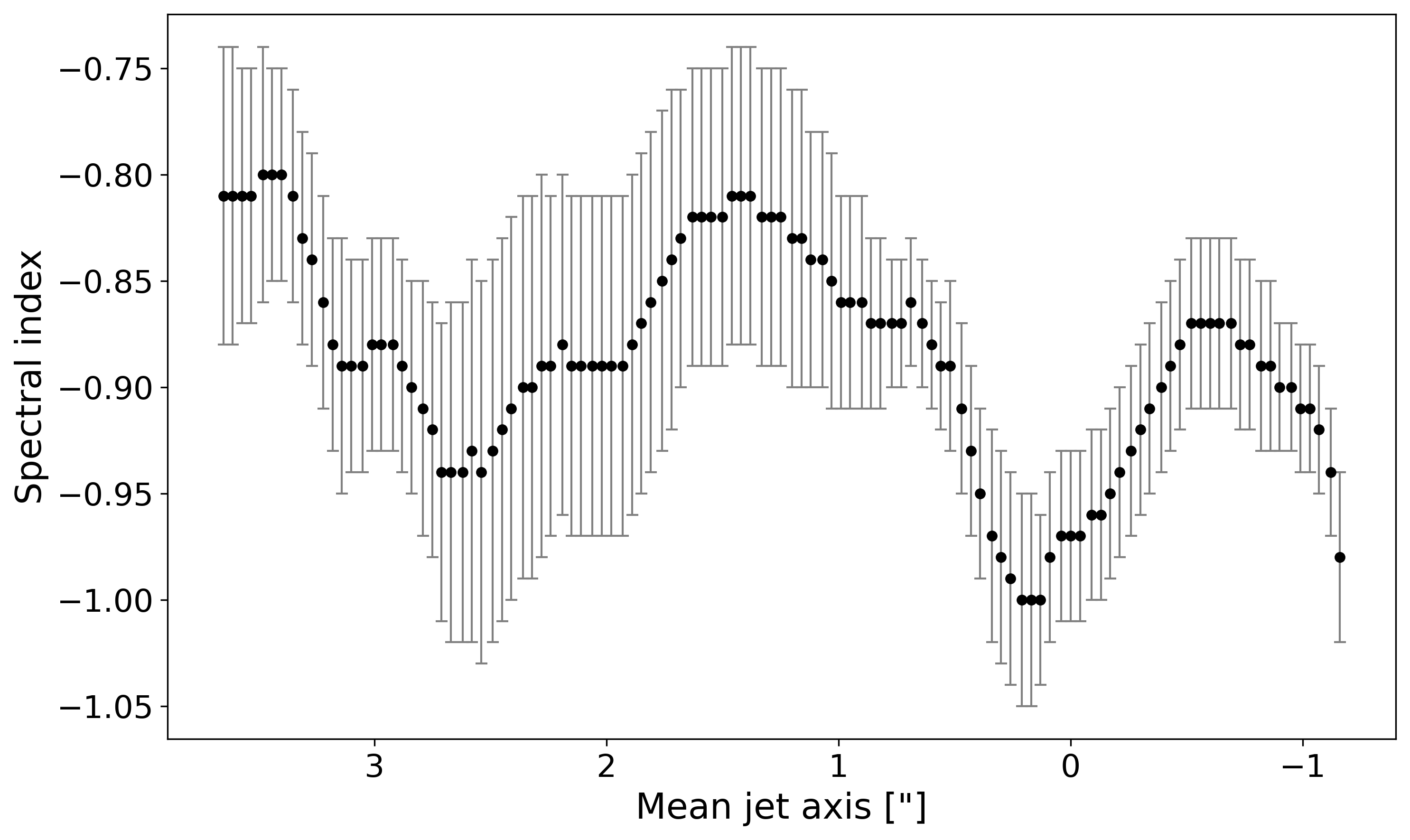}
    \caption{\textit{Top}: Spectral index map of the outer kiloparsec-scale jet between 138 and 350 GHz. The spectral index is indicated with the color bar, and the gray contours correspond to the total intensity contours at 138 GHz. The solid black line shows the total intensity ridge line at 138 GHz. The insets show the spectral index fits at specific pixel locations, indicated by the red arrows. The solid black line shows a simple power-law fit, with the slope corresponding to the color scale, and the dashed black line shows a log-parabola fit. \textit{Bottom}: Spectral index along the total intensity ridge line.}
    \label{fig:hotspot_spidx}
\end{figure}

We used the images of each Spw for all the bands smoothed to the beam size of Spw 0 in band 4 to produce a spectral index map between 138 and 350\,GHz. The spectral index was calculated by fitting a linear function to the total intensity against the frequency in log-space using all the 12 Spws. We define the spectral index sign as $I \propto \nu^{+\alpha}$. We only used pixels where the total intensity in each pixel is higher than three times the rms uncertainty of the given image. The total intensity uncertainty used in the fitting is the rms uncertainty of each image and a 10\% absolute calibration uncertainty added in quadrature. We only accepted fits with $\chi^2<18.307$, corresponding to the 95\% confidence limit for 10 degrees of freedom. The resulting spectral index map is shown in Fig.~\ref{fig:hotspot_spidx}. We also extracted a total intensity ridge line from the band 4 Spw 0 image, and extracted the spectral index values along the ridge. These are also shown in Fig.~\ref{fig:hotspot_spidx}. 

The spectral index distribution is relatively steep, around $-0.9$, but several flatter portions of $-0.8$ are seen along the ridge. The spectral index steepens at the hotspot location, which indicates that the synchrotron losses are more severe at that point. The structure we observe is very similar to what is seen in spectral index images between 15 and 22\,GHz from VLA data \citep{perley17}, including the slight asymmetries in the spectral index where the upper half has somewhat flatter indices than the bottom half. In Fig~\ref{fig:hotspot_spidx} we also show the spectral index fits in two locations of the jet. There are locations along the jet where there seems to be some curvature in the spectrum instead of it following a simple power-law form. A curved fit could indicate that synchrotron losses are more severe at the highest frequencies. We found that a log-parabola shape results in a smaller reduced $\chi^2$; however, according to the Bayesian information criterion (BIC), we cannot disfavor the simpler power-law fit.

\section{Discussion}\label{section:discussion}

\subsection{Polarization and Faraday rotation in the core}
\begin{figure}
    \centering
    \includegraphics[width=1.0\linewidth]{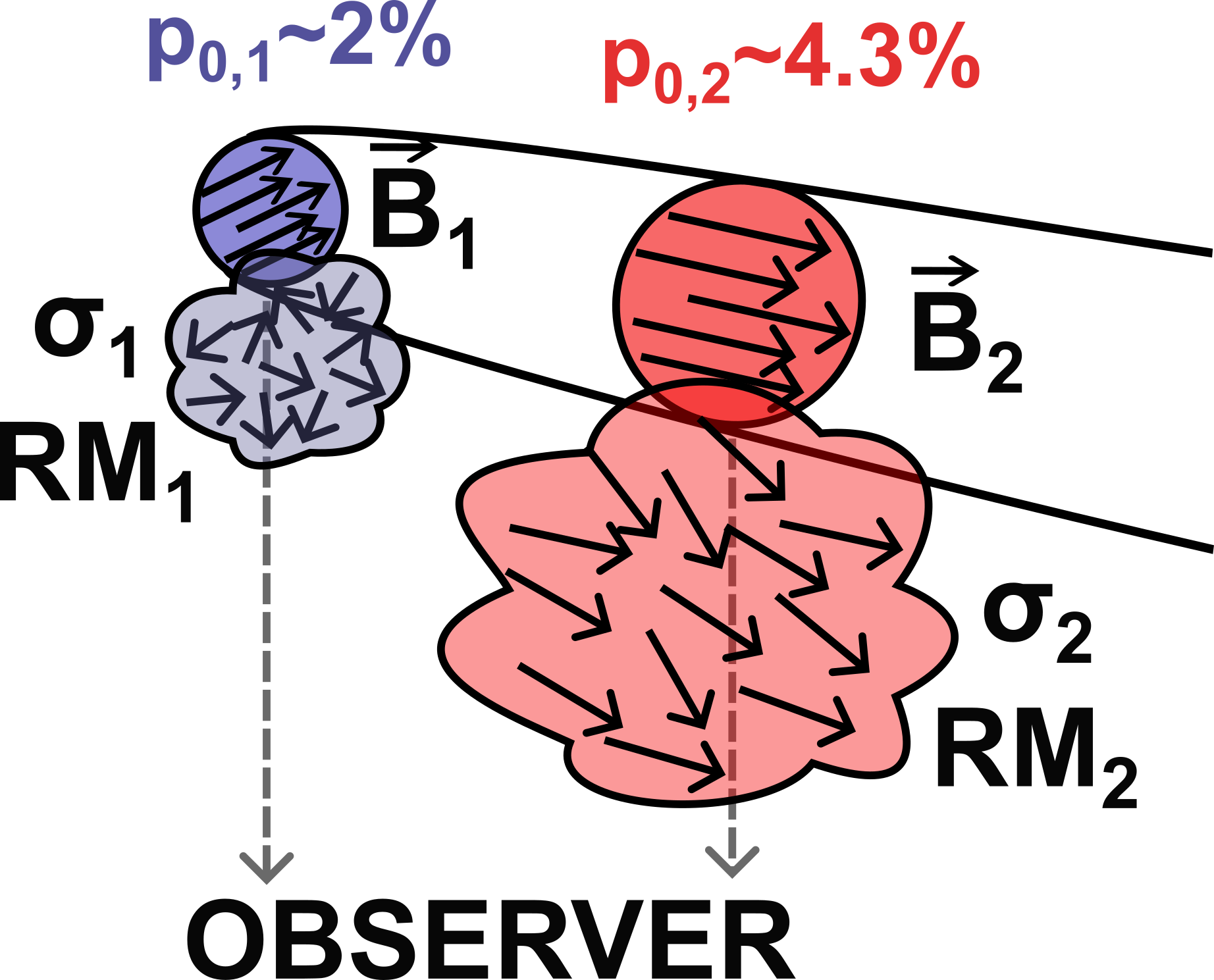}
    \caption{Simplified schematic showing the potential spatial distribution of our best-fitting T2 model of two external Faraday screens. Spatially resolved observations are required to verify the validity of our model.}
    \label{fig:sketch}
\end{figure}

The high RM ($\simeq +2.6\times10^{5}$~\si{rad.m^{-2}}) and  $\sigma_{\mathrm{RM}}$ ($\simeq 2.4\times10^{5}$~\si{rad.m^{-2}}) suggest a compact, magnetized plasma near the base of the relativistic jet acting as a Faraday screen, while the low-RM component ($\simeq +1.5\times10^{4}$~\si{rad.m^{-2}}) likely 
originates in a more extended, less turbulent region of the jet or surrounding medium. A simple schematic depicting this is shown in Fig.~\ref{fig:sketch}. These results are consistent with a stratified or multi-zone jet model for 3C~273, where multiple synchrotron-emitting regions experience different local magnetic field and density conditions. This is consistent with our findings in \citetalias{hovatta19}, where we found that either two polarized components or internal Faraday rotation were needed to explain the polarization behavior within ALMA band 6. With the much smaller wavelength coverage of those single-band observations, we were unable to distinguish between the two models unlike here where the single inverse depolarization model (T\textsubscript{inv}) corresponding to the internal Faraday rotation model of \citetalias{hovatta19} is clearly disfavored.

For the high-RM component, we find a comparable RM dispersion to the mean RM, while for the low-RM component the $2\sigma$ limit on the RM dispersion is $\sim$2 times lower than the mean RM. Both these results indicate a significant coherent magnetic field on the scale of the polarized jet emission region. This is in contrast to the general results found in parsec-scale jets at lower frequencies of 8 to 15 GHz with the Very Long Baseline Array within the MOJAVE program \citep{hovatta12}, where on average they found that the random field dominated for most sources. However, even at the MOJAVE frequencies the 3C~273 jet was an outlier in this regard, with its transverse RM gradient and inverse depolarization trend indicating the present of significantly ordered fields and possibly some internal Faraday rotation effects. 
From VLBI studies of 3C~273 up to 86 GHz \citep{attridge05}, we can consider a jet magnetic field strength of at least 6 G \citep{savolainen2008} and jet size of order 1 pc at these frequencies. Along with a thermal electron density of $6\times10^{-2}$~cm$^{-3}$ \citep{roeser00}, these parameters can provide mean RMs and dispersions of order $10^5$\rad\  that can readily account for the high-RM component properties. The low-RM component of $10^4$\rad, with much lower dispersion, suggests a uniform Faraday screen close to the jet and/or a compact emission region that probes a narrow enough line of sight to not experience significant RM fluctuations across the emission region. 

Although we analyzed the integrated emission from the core (i.e.,~all jet emission within $\sim$2 kpc), we note that the jet emission is found to be optically thin at these frequencies ($\alpha_{\rm core}=-0.7915\pm0.0003$). Therefore, we should not discount the possibility that the polarized emission of at least one of the RM components is coming from the jet base (within, e.g.,~10$r_g$). This would be similar to the two-component model proposed for M87 in \cite{goddi21}, where it was suggested that the RM and polarization at 3\,mm and 1.3\,mm observations with ALMA alone were probing two Faraday components, one very compact at the nucleus seen also by the Event Horizon Telescope (EHT) and another more extended and less variable one that was seen only in the ALMA-only observations probing larger scales.

The EHT polarization observations of M87* \citep{ehtM87pol} have provided direct constraints on the plasma properties in the jet-launching region on such scales. In a simple one-zone model of the polarized 230 GHz emission, they inferred characteristic electron densities $n_e\sim10^4-10^7$~cm$^{-3}$ and magnetic field strengths of $B\sim1-30$~G within a few gravitational radii. 
While the global disk structures differ substantially (M87* accretes via a radiatively inefficient flow, whereas 3C~273 is a luminous quasar with a radiatively efficient disk), we can do a simple comparison if we focus on the EHT M87* results  within the jet-launching region where the magnetization is greater than one. 
In M87* the relevant parameters in this regime were found to be closer to $n_e\sim10^4$~cm$^{-3}$ and $B\sim30$~G. We can estimate the resulting rotation measure using
\begin{equation}
    RM = 0.81 \int n_e \mathbf{B} \cdot \mathbf{\mathrm{d}l} ~\mathrm{rad/m}^2,
\end{equation}
where $n_e$ is given in cm$^{-3}$, $B$ in $\mu$G, and the path along the line of sight d$\mathbf{l}$ in parsecs. Adopting the latest black hole mass estimate for 3C~273, $M_\mathrm{BH} = 10^{9.1}$\,M$_\odot$ by \cite{li22} gives us the scale $r_g = 6\times10^{-5}$\,pc. Assuming a path length of 10$r_g$ and the parameters for M87* would result in RMs on the order of $10^8$\rad, which is higher than we observe. 

However, the magnetic field strength of the highly accreting 3C~273 can be even higher than in M87*.
Similar to \citetalias{hovatta19}, we can estimate the magnetic field strength by assuming that the magnetic field follows the relation $B\propto r^{-1}$ as expected for the toroidal component of the field. \cite{savolainen2008} estimated the magnetic field strength of 3C~273 to be $\sim2$~G at a distance of $\leq0.06$~mas from the jet apex using multifrequency VLBI observations. Assuming an emission region at 10$r_g$, would give us a magnetic field strength of $\sim75$\,G, which would result in even higher RM if the electron density remains the same. 

On the other hand, the magnetic field we estimate is within the jet itself, which can be much higher than in any external screens. We also do not know the path length along the line of sight. Therefore, in order to produce RMs of $10^5$ to $10^6$\rad, for the magnetic field strength of 75~G we would need a substantially lower electron density of $n_e\sim10-100$~cm$^{-3}$ than in M87*, or alternatively the Faraday screen to contain a much lower magnetic field of $\lesssim1$~G. Of course it is also possible that the dominating polarized regions are much further down the jet, but the scenario here should not be discounted as it readily explains the properties of the high-RM component, and the observed RM time variability. 
This makes 3C~273 an excellent candidate for polarization imaging with the EHT, to better constrain the origin and nature of the polarized emission and Faraday rotating gas in a distinctly different accreting regime than M87*. 

\subsection{Variability in the core RM}
\begin{figure}
    \includegraphics[width=1.0\linewidth]{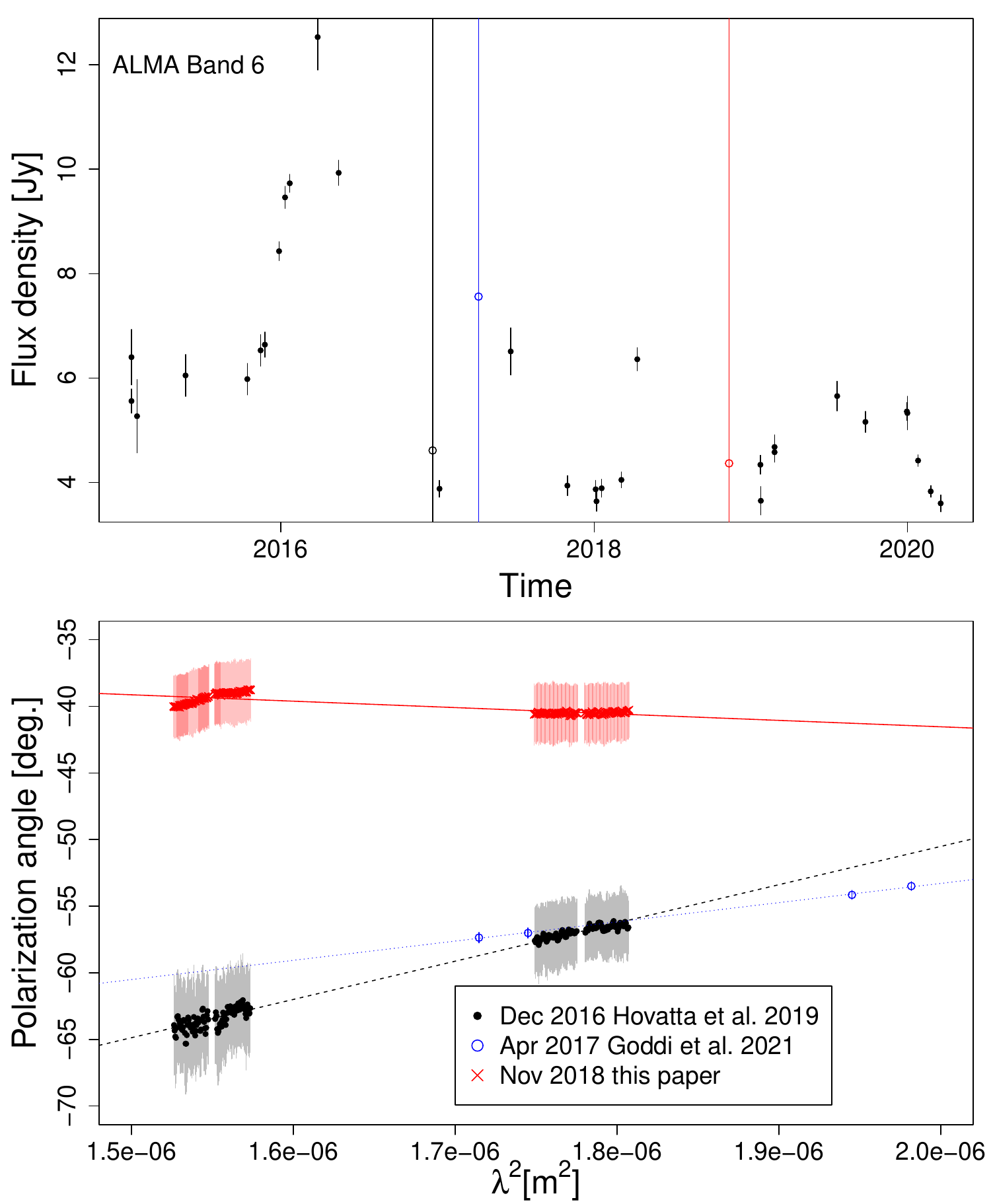}
    \caption{\textit{Top}: Light curve of 3C~273 in ALMA band 6 (233 GHz) from the ALMA Calibrator Source Catalogue. Colored vertical lines indicate the epochs of different ALMA full-polarization observations; open symbols mark the mean total intensity values from those observations. \textit{Bottom}: Polarization angle against wavelength-squared for the three different epochs along with simple linear fits indicating the amount of RM when assuming a single external Faraday-rotating component.}
    \label{fig:flux_coreRM}
\end{figure}

We can also directly compare our band 6 results with \citetalias{hovatta19}. Figure~\ref{fig:flux_coreRM} shows the EVPA measurements from \citetalias{hovatta19} and this paper along with the data from \cite{goddi21} that were taken during the EHT campaign in April 2017. The RM value of $\sim2.5\times10^{5}$~\si{rad.m^{-2}} obtained by \cite{goddi21} is smaller than the value of $\sim5.0\times10^{5}$~\si{rad.m^{-2}} in \citetalias{hovatta19} even though there is only four months between the observations. Our new data taken over 1.5 years later show very different behavior with a nearly flat polarization angle over the 1.3 mm band, resulting in a much smaller nominal RM of $\sim-8.4\times10^4$~\si{rad.m^{-2}} if a simple linear fit as shown in Fig.~\ref{fig:flux_coreRM} is performed. This is similar to the variability seen in M87 on timescales of a year \citep{goddi21}.

At the same time, the total intensity of this highly variable source has also changed between the different epochs as shown in the top panel of Fig.~\ref{fig:flux_coreRM} where we show the flux density evolution of the source from the ALMA Calibrator Source Catalogue\footnote{\url{https://almascience.eso.org/sc/}}. The variability is associated with the beamed emission from the core, which can be connected with the emergence of new components in the parsec-scale jet \citep[e.g.,][]{savolainen02}. These are blended within the beam of the ALMA-only observations, resulting in complex polarization behavior across the bands as observed here in our multiband observations. Spatially resolved millimeter-wavelength observations of the core are required to further probe the Faraday rotation structure and the number of polarized components in the source.

\subsection{Polarization of the kiloparsec-scale jet}
The polarization structure of the kiloparsec-scale jet in our ALMA millimeter-wavelength observations is very similar to what has been seen at centimeter wavelengths \citep[e.g.,][]{conway93, perley17}. In all the ALMA bands the brightest hotspot H2 where our images are centered is highly polarized with a fractional polarization of 20-30\%. This is similar to \cite{conway93} who found that at 6\,cm H2 is highly polarized with 29\% fractional polarization and at 18cm it is 23\% with a comparable resolution of $0.4\arcsec$ to our 145\,GHz data. The hotspot H3 located about $1\arcsec$ closer to the nucleus has a lower polarization, which \cite{conway93} suggest is due to line-of-sight or beam smearing of two oppositely polarized components. Overall, the variations we observe along the ridge line look very similar to those at centimeter wavelengths \citep{conway93}, suggesting that the same population of synchrotron-emitting electrons is also responsible for the millimeter-wavelength emission.

\cite{conway93} detected a very low Faraday rotation of around 2~\si{rad.m^{-2}} in the jet with a transverse RM gradient of 0.8~\si{rad.m^{-2}} across the jet. 
We attempted to make an RM map of the kiloparsec-scale jet by using the data of the individual Spws similar to the spectral index map, but due to the relatively poor sensitivity and larger EVPA uncertainties, we do not detect any RM deviating from 0~\si{rad.m^{-2}} within three times the RM uncertainty. However, due to the small wavelength-squared coverage in our data, we  are not sensitive to the small values seen in the centimeter-wavelength data.

The magnetic field direction is predominantly poloidal along the jet direction until about $2\arcsec$ from the brightest hotspot H2. Closer to the hotspot, the magnetic field in the inner jet is perpendicular to the jet, while around the edges the magnetic field follows the jet direction. Such a circumferential magnetic field structure around the jet edges can occur simply due to compression and shearing of an originally tangled magnetic field \citep{laing80}. This kind of polarization structure has been observed in several other quasars at centimeter-wavelengths \citep[e.g.,][]{bridle94}.

Early numerical simulations by \cite{matthews90} show a very similar magnetic field structure, where an initially tangled magnetic field is compressed at the strong shock resulting in a perpendicular magnetic field at the shock, while significant shearing between the jet and the cocoon backflow results in the magnetic field to follow the jet edges. The polarization is expected to be the highest at the jet edges, as is seen in our data, although \cite{matthews90} point out that the fractional polarization in their simulations is much higher than typically observed, near 70\% at the jet edges. 

Interpretation of the circumferential magnetic field direction is complicated, though, because also a toroidal magnetic field will result in a similar structure due to projection effects for jets viewed at small angles to the line of sight \citep{clarke89}. We see a clear asymmetry between the total intensity and fractional polarization peaks across the jet at the hotspot location in Fig.~\ref{fig:slices} where the polarization peak is shifted to the lower side of the jet. In beamed parsec-scale jets, these kinds of transverse asymmetries can be related to large-scale helical magnetic fields \citep{clausen-brown11}, and this has been seen in the parsec-scale jet of 3C~273 in VLBI observations \citep{hovatta12}. Given that there is no apparent bending between the parsec- and kiloparsec-scale jets of 3C~273, we can assume that the viewing angles are similar \citep[see also][]{meyer16}. However, it is difficult to distinguish this from effects due to smearing of two oppositely polarized components. 

More recent 3D magnetohydrodynamic simulations confirm how stretching an initially tangled magnetic field along the jet creates a poloidally dominated magnetic field structure with shearing at the jet edges \citep{huarte11}. Similar structures are also seen in very recent 3D relativistic magnetohydrodynamic simulations that account for the synchrotron and inverse Compton losses in post-processing using an evolving electron energy distribution that also accounts for particle (re)acceleration \citep{jerrim25}. These simulations also show circumferential magnetic fields around the jet edges, especially in the simulations with stronger magnetic fields, although the polarization structure in the jet itself is patchier than what is seen in our observations of 3C~273.

\subsection{Multiwavelength connection}
The kiloparsec-scale jet of 3C~273 has been studied in multiple wavelengths ever since its discovery in the 1960s \citep{schmidt1963, greenstein64}. High-energy X-ray emission from it was discovered early on by the \textit{Einstein} Observatory \citep{willingale81} with the first \textit{Chandra} observations revealing the detailed X-ray emission structure \citep{sambruna01}. Modeling of the spectrum of the different knots from radio to X-ray energies has shown that the emission processes are complex with the radio and near-infrared radiation originating from a different emission component than the optical and X-ray radiation \citep{uchiyama06}. Especially the origin of the X-ray emission is unclear as the most traditionally assumed assumption that it is inverse Compton scattering off the cosmic microwave background photons is ruled out for 3C~273 based on the upper limit of its GeV emission by the {\it Fermi}-LAT instrument \citep{meyer14}. Alternatively, the X-ray emission could be synchrotron emission from another population of electrons \citep{uchiyama06}.

Such a two-component electron population would require two separate acceleration regions, such as shocks and a shear layer around the jet, as proposed by \cite{stawarz02}. 
Recently, the Very Energetic Radiation Imaging Telescope Array System (VERITAS) reported the discovery of very-high-energy gamma-ray emission at energies $> 350$~GeV from the direction of 3C~273\footnote{Presented at the 2025 ICRC conference: \url{https://indico.cern.ch/event/1258933/contributions/6491204/attachments/3104111/5500883/Benbow_ICRC2025.pdf}}. Although VERITAS cannot spatially resolve the kiloparsec-scale jet, the gamma-ray spectrum they obtain is incompatible with the lower GeV-energy spectrum from the {\it Fermi}-LAT. Their preliminary analysis suggests that the emission is inverse Compton emission from the kiloparsec-scale jet. This would fit with the model of \cite{stawarz02} who predict that very-high-energy gamma-ray emission could be generated by the second population of synchrotron-emitting electrons in the jet.

Indeed, following the VERITAS discovery, \cite{tavecchio25} proposed a model where the kiloparsec-scale jet at knots A and B contains two electron populations, one responsible for the radio and optical emission, and another for the very-high-energy gamma-ray emission. The low-energy electrons would be accelerated via the standard diffusive shock acceleration in the knots, while the high-energy electron component would consist of electrons escaping the shocked region into a sheath where they would undergo further shear acceleration. Both components would emit synchrotron and inverse Compton emission, and the high-energy electrons scattering off cosmic microwave background photons would be responsible for the very-high-energy gamma-ray emission detected by VERITAS. 

In this model, the very-high-energy gamma-ray emission would originate in the brightest X-ray knots, A and B, but unfortunately the sensitivity of our polarization maps is not high enough to detect polarization across the jet in this region to study if it supports the scenario of a sheared layer around the jet, as seen in the hotspot region. The X-ray emission of the hotspot region is very faint but even in the hotspot region the optical-ultraviolet emission show a flattening of the spectrum, which could hint at two electron populations \citep{uchiyama06}. Detailed comparison of high-resolution observations of the millimeter and optical polarization in the hotspot region, similar to \cite{cara13}, may shed light on this but is beyond the scope of the current paper.

\section{Conclusions}\label{conclusions}
We observed 3C~273 with ALMA in three different wavelengths, around 2, 1.3, and 0.8\,mm, in November 2018. Our observations were motivated by our previous study \citepalias{hovatta19}, in which very large in-band Faraday rotation was detected over the 1.3\,mm band in the nucleus of the source but we could not discriminate between different Faraday rotation models due to the narrow wavelength-squared coverage. In our multiband observations we targeted both the nucleus and the kiloparsec-scale jet of 3C~273.

The data of the nucleus are best modeled with two polarized components with external Faraday dispersion where one has a very high RM of $+2.6\times10^{5}$~\si{rad.m^{-2}} and the other a moderate RM of $+1.5\times10^{4}$~\si{rad.m^{-2}}. The former likely originates from a compact magnetized region near the base of the jet, while the latter may originate from a more extended region of the jet or the surrounding medium. Our multiband data clearly disfavor models with a single polarized component. We detect variability in the polarization angle on timescales of a week, which further suggests that the emission of the nucleus is connected with the parsec-scale beamed jet. Spatially resolved millimeter-wavelength observations with, for example, the EHT are needed to map the polarized components and confirm our findings.

The total intensity and polarization structure of the kiloparsec-scale jet is very similar to earlier observations in the centimeter wavelengths \citep[e.g.,][]{conway93,perley17}. The magnetic field is seen to be predominantly along the jet direction until about 2\arcsec\ from the brightest hotspot region, where in the inner jet region the magnetic field turns perpendicular to the jet, indicative of a transverse shock. The magnetic field direction in the hotspot region follows the jet edges, which can result from the shearing of an originally tangled magnetic field \citep{laing80}. The similar polarization structure in centimeter and millimeter wavelengths suggests that the same electron population is responsible for their emission, although more detailed modeling that compares the optical and millimeter-wavelength polarization would be needed to confirm this.  

\section*{Data availability}
The polarization data used in the QU-modeling are available as a table poldata.dat in electronic form at the CDS via \url{http://cdsweb.u-strasbg.fr/cgi-bin/qcat?J/A+A/}. The Stokes I, Q, and U maps of the outer kiloparsec-scale jet are available in fits format at the CDS. 

\begin{acknowledgements}
 We thank the referee, Dr. Russel Taylor, for the comments that improved the clarity of the paper. TH acknowledges support from the Research Council of Finland projects 317383, 320085, 345899, and 362571 and from the European Union ERC-2024-COG - PARTICLES - 101169986. Views and opinions expressed are however those of the author(s) only and do not necessarily reflect those of the European Union or the European Research Council Executive Agency. Neither the European Union nor the granting authority can be held responsible for them. 
SPO acknowledges support from the Comunidad de Madrid Atracción de Talento program via grant 2022-T1/TIC-23797, and grant PID2023-146372OB-I00 funded by MICIU/AEI/10.13039/501100011033 and by ERDF, EU. TS acknowledges support from the Research Council of Finland projects 362572 and 365088.
This paper makes use of the following ALMA data: ADS/JAO.ALMA\#2018.1.00357.S. ALMA is a partnership of ESO (representing its member states), NSF (USA) and NINS (Japan), together with NRC (Canada), NSTC and ASIAA (Taiwan), and KASI (Republic of Korea), in cooperation with the Republic of Chile. The Joint ALMA Observatory is operated by ESO, AUI/NRAO and NAOJ. The National Radio Astronomy Observatory is a facility of the National Science Foundation operated under cooperative agreement by Associated Universities, Inc.
\end{acknowledgements}

\bibliographystyle{aa} 
\bibliography{references.bib} 

@ARTICLE{ehtM87pol,
       author = {{Event Horizon Telescope Collaboration} and {Akiyama}, Kazunori and {Algaba}, Juan Carlos and {Alberdi}, Antxon and {Alef}, Walter and {Anantua}, Richard and {Asada}, Keiichi and {Azulay}, Rebecca and {Baczko}, Anne-Kathrin and {Ball}, David and {Balokovi{\'c}}, Mislav and {Barrett}, John and {Benson}, Bradford A. and {Bintley}, Dan and {Blackburn}, Lindy and {Blundell}, Raymond and {Boland}, Wilfred and {Bouman}, Katherine L. and {Bower}, Geoffrey C. and {Boyce}, Hope and {Bremer}, Michael and {Brinkerink}, Christiaan D. and {Brissenden}, Roger and {Britzen}, Silke and {Broderick}, Avery E. and {Broguiere}, Dominique and {Bronzwaer}, Thomas and {Byun}, Do-Young and {Carlstrom}, John E. and {Chael}, Andrew and {Chan}, Chi-kwan and {Chatterjee}, Shami and {Chatterjee}, Koushik and {Chen}, Ming-Tang and {Chen}, Yongjun and {Chesler}, Paul M. and {Cho}, Ilje and {Christian}, Pierre and {Conway}, John E. and {Cordes}, James M. and {Crawford}, Thomas M. and {Crew}, Geoffrey B. and {Cruz-Osorio}, Alejandro and {Cui}, Yuzhu and {Davelaar}, Jordy and {De Laurentis}, Mariafelicia and {Deane}, Roger and {Dempsey}, Jessica and {Desvignes}, Gregory and {Dexter}, Jason and {Doeleman}, Sheperd S. and {Eatough}, Ralph P. and {Falcke}, Heino and {Farah}, Joseph and {Fish}, Vincent L. and {Fomalont}, Ed and {Ford}, H. Alyson and {Fraga-Encinas}, Raquel and {Friberg}, Per and {Fromm}, Christian M. and {Fuentes}, Antonio and {Galison}, Peter and {Gammie}, Charles F. and {Garc{\'\i}a}, Roberto and {Gelles}, Zachary and {Gentaz}, Olivier and {Georgiev}, Boris and {Goddi}, Ciriaco and {Gold}, Roman and {G{\'o}mez}, Jos{\'e} L. and {G{\'o}mez-Ruiz}, Arturo I. and {Gu}, Minfeng and {Gurwell}, Mark and {Hada}, Kazuhiro and {Haggard}, Daryl and {Hecht}, Michael H. and {Hesper}, Ronald and {Himwich}, Elizabeth and {Ho}, Luis C. and {Ho}, Paul and {Honma}, Mareki and {Huang}, Chih-Wei L. and {Huang}, Lei and {Hughes}, David H. and {Ikeda}, Shiro and {Inoue}, Makoto and {Issaoun}, Sara and {James}, David J. and {Jannuzi}, Buell T. and {Janssen}, Michael and {Jeter}, Britton and {Jiang}, Wu and {Jimenez-Rosales}, Alejandra and {Johnson}, Michael D. and {Jorstad}, Svetlana and {Jung}, Taehyun and {Karami}, Mansour and {Karuppusamy}, Ramesh and {Kawashima}, Tomohisa and {Keating}, Garrett K. and {Kettenis}, Mark and {Kim}, Dong-Jin and {Kim}, Jae-Young and {Kim}, Jongsoo and {Kim}, Junhan and {Kino}, Motoki and {Koay}, Jun Yi and {Kofuji}, Yutaro and {Koch}, Patrick M. and {Koyama}, Shoko and {Kramer}, Michael and {Kramer}, Carsten and {Krichbaum}, Thomas P. and {Kuo}, Cheng-Yu and {Lauer}, Tod R. and {Lee}, Sang-Sung and {Levis}, Aviad and {Li}, Yan-Rong and {Li}, Zhiyuan and {Lindqvist}, Michael and {Lico}, Rocco and {Lindahl}, Greg and {Liu}, Jun and {Liu}, Kuo and {Liuzzo}, Elisabetta and {Lo}, Wen-Ping and {Lobanov}, Andrei P. and {Loinard}, Laurent and {Lonsdale}, Colin and {Lu}, Ru-Sen and {MacDonald}, Nicholas R. and {Mao}, Jirong and {Marchili}, Nicola and {Markoff}, Sera and {Marrone}, Daniel P. and {Marscher}, Alan P. and {Mart{\'\i}-Vidal}, Iv{\'a}n and {Matsushita}, Satoki and {Matthews}, Lynn D. and {Medeiros}, Lia and {Menten}, Karl M. and {Mizuno}, Izumi and {Mizuno}, Yosuke and {Moran}, James M. and {Moriyama}, Kotaro and {Moscibrodzka}, Monika and {M{\"u}ller}, Cornelia and {Musoke}, Gibwa and {Mus Mej{\'\i}as}, Alejandro and {Michalik}, Daniel and {Nadolski}, Andrew and {Nagai}, Hiroshi and {Nagar}, Neil M. and {Nakamura}, Masanori and {Narayan}, Ramesh and {Narayanan}, Gopal and {Natarajan}, Iniyan and {Nathanail}, Antonios and {Neilsen}, Joey and {Neri}, Roberto and {Ni}, Chunchong and {Noutsos}, Aristeidis and {Nowak}, Michael A. and {Okino}, Hiroki and {Olivares}, H{\'e}ctor and {Ortiz-Le{\'o}n}, Gisela N. and {Oyama}, Tomoaki and {{\"O}zel}, Feryal and {Palumbo}, Daniel C.~M. and {Park}, Jongho and {Patel}, Nimesh and {Pen}, Ue-Li and {Pesce}, Dominic W. and {Pi{\'e}tu}, Vincent and {Plambeck}, Richard and {PopStefanija}, Aleksandar and {Porth}, Oliver and {P{\"o}tzl}, Felix M. and {Prather}, Ben and {Preciado-L{\'o}pez}, Jorge A. and {Psaltis}, Dimitrios and {Pu}, Hung-Yi and {Ramakrishnan}, Venkatessh and {Rao}, Ramprasad and {Rawlings}, Mark G. and {Raymond}, Alexander W. and {Rezzolla}, Luciano and {Ricarte}, Angelo and {Ripperda}, Bart and {Roelofs}, Freek and {Rogers}, Alan and {Ros}, Eduardo and {Rose}, Mel and {Roshanineshat}, Arash and {Rottmann}, Helge and {Roy}, Alan L. and {Ruszczyk}, Chet and {Rygl}, Kazi L.~J. and {S{\'a}nchez}, Salvador and {S{\'a}nchez-Arguelles}, David},
        title = "{First M87 Event Horizon Telescope Results. VIII. Magnetic Field Structure near The Event Horizon}",
      journal = {\apjl},
         year = 2021,
        month = mar,
       volume = {910},
       number = {1},
          eid = {L13},
        pages = {L13},
          doi = {10.3847/2041-8213/abe4de},
archivePrefix = {arXiv},
       eprint = {2105.01173},
 primaryClass = {astro-ph.HE},
       adsurl = {https://ui.adsabs.harvard.edu/abs/2021ApJ...910L..13E}
}

@inproceedings{savolainen2008,
	Adsurl = {http://adsabs.harvard.edu/abs/2008ASPC..386..451S},
	Archiveprefix = {arXiv},
	Author = {{Savolainen}, T. and {Wiik}, K. and {Valtaoja}, E. and {Tornikoski}, M.},
	Booktitle = {Extragalactic Jets: Theory and Observation from Radio to Gamma Ray},
	Editor = {{Rector}, T.~A. and {De Young}, D.~S.},
	Eprint = {0708.0144},
	Month = jun,
	Pages = {451},
	Series = {Astronomical Society of the Pacific Conference Series},
	Title = {{Magnetic Field Structure in the Parsec Scale Jet of 3C 273 from Multifrequency VLBA Observations}},
	Volume = 386,
	Year = 2008}

@article{asada02,
	Adsurl = {http://adsabs.harvard.edu/abs/2002PASJ...54L..39A},
	Author = {{Asada}, K. and {Inoue}, M. and {Uchida}, Y. and {Kameno}, S. and {Fujisawa}, K. and {Iguchi}, S. and {Mutoh}, M.},
	Doi = {10.1093/pasj/54.3.L39},
	Eprint = {astro-ph/0205497},
	Journal = {\pasj},
	Month = jun,
	Pages = {L39-L43},
	Title = {{A Helical Magnetic Field in the Jet of 3C 273}},
	Volume = 54,
	Year = 2002}

@article{attridge05,
	Adsurl = {http://adsabs.harvard.edu/abs/2005ApJ...633L..85A},
	Author = {{Attridge}, J.~M. and {Wardle}, J.~F.~C. and {Homan}, D.~C.},
	Doi = {10.1086/498392},
	Eprint = {astro-ph/0506243},
	Journal = {\apjl},
	Month = nov,
	Pages = {L85-L88},
	Title = {{Concurrent 43 and 86 GHz Very Long Baseline Polarimetry of 3C 273}},
	Volume = 633,
	Year = 2005}

@ARTICLE{bahcall95,
       author = {{Bahcall}, J.~N. and {Kirhakos}, S. and {Schneider}, D.~P. and {Davis}, R.~J. and {Muxlow}, T.~W.~B. and {Garrington}, S.~T. and {Conway}, R.~G. and {Unwin}, S.~C.},
        title = "{Hubble Space Telescope and MERLIN Observations of the Jet in 3C 273}",
      journal = {\apjl},
         year = 1995,
        month = oct,
       volume = {452},
        pages = {L91},
          doi = {10.1086/309717},
archivePrefix = {arXiv},
       eprint = {astro-ph/9509028},
 primaryClass = {astro-ph},
       adsurl = {https://ui.adsabs.harvard.edu/abs/1995ApJ...452L..91B}
}

@ARTICLE{bridle94,
       author = {{Bridle}, Alan H. and {Hough}, David H. and {Lonsdale}, Colin J. and {Burns}, Jack O. and {Laing}, Robert A.},
        title = "{Deep VLA Imaging of Twelve Extended 3CR Quasars}",
      journal = {\aj},
         year = 1994,
        month = sep,
       volume = {108},
        pages = {766},
          doi = {10.1086/117112},
       adsurl = {https://ui.adsabs.harvard.edu/abs/1994AJ....108..766B}
}

@article{cara13,
doi = {10.1088/0004-637X/773/2/186},
url = {https://doi.org/10.1088/0004-637X/773/2/186},
year = {2013},
month = {aug},
publisher = {The American Astronomical Society},
volume = {773},
number = {2},
pages = {186},
author = {Cara, Mihai and Perlman, Eric S. and Uchiyama, Yasunobu and Cheung, Chi C. and Coppi, Paolo S. and Georganopoulos, Markos and Worrall, Diana M. and Birkinshaw, Mark and Sparks, William B. and Marshall, Herman L. and Stawarz, Lukasz and Begelman, Mitchell C. and O'Dea, Christopher P. and Baum, Stefi A.},
title = {POLARIMETRY AND THE HIGH-ENERGY EMISSION MECHANISMS IN QUASAR JETS: THE CASE OF PKS 1136−135},
journal = {The Astrophysical Journal}
}

@ARTICLE{clarke89,
       author = {{Clarke}, David A. and {Norman}, Michael L. and {Burns}, Jack O.},
        title = "{Numerical Observations of a Simulated Jet with a Passive Helical Magnetic Field}",
      journal = {\apj},
         year = 1989,
        month = jul,
       volume = {342},
        pages = {700},
          doi = {10.1086/167631},
       adsurl = {https://ui.adsabs.harvard.edu/abs/1989ApJ...342..700C}
}

@ARTICLE{clausen-brown11,
       author = {{Clausen-Brown}, E. and {Lyutikov}, M. and {Kharb}, P.},
        title = "{Signatures of large-scale magnetic fields in active galactic nuclei jets: transverse asymmetries}",
      journal = {\mnras},
         year = 2011,
        month = aug,
       volume = {415},
       number = {3},
        pages = {2081-2092},
          doi = {10.1111/j.1365-2966.2011.18757.x},
archivePrefix = {arXiv},
       eprint = {1101.5149},
 primaryClass = {astro-ph.HE},
       adsurl = {https://ui.adsabs.harvard.edu/abs/2011MNRAS.415.2081C}
}

@ARTICLE{conway93,
       author = {{Conway}, R.~G. and {Garrington}, S.~T. and {Perley}, R.~A. and {Biretta}, J.~A.},
        title = "{Synchrotron radiation from the jet of 3C 273. II. The radio structure and polarization.}",
      journal = {\aap},
         year = 1993,
        month = jan,
       volume = {267},
        pages = {347-362},
       adsurl = {https://ui.adsabs.harvard.edu/abs/1993A&A...267..347C}
}

@ARTICLE{conway94,
       author = {{Conway}, R.~G. and {Davis}, R.~J.},
        title = "{Synchrotron radiation from the jet of 3C 273 III. The speed and direction of the jet.}",
      journal = {\aap},
         year = 1994,
        month = apr,
       volume = {284},
        pages = {724-730},
       adsurl = {https://ui.adsabs.harvard.edu/abs/1994A&A...284..724C}
}

@ARTICLE{goddi19,
       author = {{Goddi}, C. and {Mart{\'\i}-Vidal}, I. and {Messias}, H. and {Crew}, G.~B. and {Herrero-Illana}, R. and {Impellizzeri}, V. and {Rottmann}, H. and {Wagner}, J. and {Fomalont}, E. and {Matthews}, L.~D. and {Petry}, D. and {Phillips}, N. and {Tilanus}, R. and {Villard}, E. and {Blackburn}, L. and {Janssen}, M. and {Wielgus}, M.},
        title = "{Calibration of ALMA as a Phased Array. ALMA Observations During the 2017 VLBI Campaign}",
      journal = {\pasp},
         year = 2019,
        month = jul,
       volume = {131},
       number = {1001},
        pages = {075003},
          doi = {10.1088/1538-3873/ab136a},
archivePrefix = {arXiv},
       eprint = {1901.09987},
 primaryClass = {astro-ph.IM},
       adsurl = {https://ui.adsabs.harvard.edu/abs/2019PASP..131g5003G}
}

@ARTICLE{goddi21,
       author = {{Goddi}, Ciriaco and {Mart{\'\i}-Vidal}, Iv{\'a}n and {Messias}, Hugo and {Bower}, Geoffrey C. and {Broderick}, Avery E. and {Dexter}, Jason and {Marrone}, Daniel P. and {Moscibrodzka}, Monika and {Nagai}, Hiroshi and {Algaba}, Juan Carlos and {Asada}, Keiichi and {Crew}, Geoffrey B. and {G{\'o}mez}, Jos{\'e} L. and {Impellizzeri}, C.~M. Violette and {Janssen}, Michael and {Kadler}, Matthias and {Krichbaum}, Thomas P. and {Lico}, Rocco and {Matthews}, Lynn D. and {Nathanail}, Antonios and {Ricarte}, Angelo and {Ros}, Eduardo and {Younsi}, Ziri and {Akiyama}, Kazunori and {Alberdi}, Antxon and {Alef}, Walter and {Anantua}, Richard and {Azulay}, Rebecca and {Baczko}, Anne-Kathrin and {Ball}, David and {Balokovi{\'c}}, Mislav and {Barrett}, John and {Benson}, Bradford A. and {Bintley}, Dan and {Blackburn}, Lindy and {Blundell}, Raymond and {Boland}, Wilfred and {Bouman}, Katherine L. and {Boyce}, Hope and {Bremer}, Michael and {Brinkerink}, Christiaan D. and {Brissenden}, Roger and {Britzen}, Silke and {Broguiere}, Dominique and {Bronzwaer}, Thomas and {Byun}, Do-Young and {Carlstrom}, John E. and {Chael}, Andrew and {Chan}, Chi-kwan and {Chatterjee}, Shami and {Chatterjee}, Koushik and {Chen}, Ming-Tang and {Chen}, Yongjun and {Chesler}, Paul M. and {Cho}, Ilje and {Christian}, Pierre and {Conway}, John E. and {Cordes}, James M. and {Crawford}, Thomas M. and {Cruz-Osorio}, Alejandro and {Cui}, Yuzhu and {Davelaar}, Jordy and {De Laurentis}, Mariafelicia and {Deane}, Roger and {Dempsey}, Jessica and {Desvignes}, Gregory and {Doeleman}, Sheperd S. and {Eatough}, Ralph P. and {Falcke}, Heino and {Farah}, Joseph and {Fish}, Vincent L. and {Fomalont}, Ed and {Ford}, H. Alyson and {Fraga-Encinas}, Raquel and {Freeman}, William T. and {Friberg}, Per and {Fromm}, Christian M. and {Fuentes}, Antonio and {Galison}, Peter and {Gammie}, Charles F. and {Garc{\'\i}a}, Roberto and {Gentaz}, Olivier and {Georgiev}, Boris and {Gold}, Roman and {G{\'o}mez-Ruiz}, Arturo I. and {Gu}, Minfeng and {Gurwell}, Mark and {Hada}, Kazuhiro and {Haggard}, Daryl and {Hecht}, Michael H. and {Hesper}, Ronald and {Ho}, Luis C. and {Ho}, Paul and {Honma}, Mareki and {Huang}, Chih-Wei L. and {Huang}, Lei and {Hughes}, David H. and {Inoue}, Makoto and {Issaoun}, Sara and {James}, David J. and {Jannuzi}, Buell T. and {Jeter}, Britton and {Jiang}, Wu and {Jimenez-Rosales}, Alejandra and {Johnson}, Michael D. and {Jorstad}, Svetlana and {Jung}, Taehyun and {Karami}, Mansour and {Karuppusamy}, Ramesh and {Kawashima}, Tomohisa and {Keating}, Garrett K. and {Kettenis}, Mark and {Kim}, Dong-Jin and {Kim}, Jae-Young and {Kim}, Jongsoo and {Kim}, Junhan and {Kino}, Motoki and {Koay}, Jun Yi and {Kofuji}, Yutaro and {Koch}, Patrick M. and {Koyama}, Shoko and {Kramer}, Michael and {Kramer}, Carsten and {Kuo}, Cheng-Yu and {Lauer}, Tod R. and {Lee}, Sang-Sung and {Levis}, Aviad and {Li}, Yan-Rong and {Li}, Zhiyuan and {Lindqvist}, Michael and {Lindahl}, Greg and {Liu}, Jun and {Liu}, Kuo and {Liuzzo}, Elisabetta and {Lo}, Wen-Ping and {Lobanov}, Andrei P. and {Loinard}, Laurent and {Lonsdale}, Colin and {Lu}, Ru-Sen and {MacDonald}, Nicholas R. and {Mao}, Jirong and {Marchili}, Nicola and {Markoff}, Sera and {Marscher}, Alan P. and {Matsushita}, Satoki and {Medeiros}, Lia and {Menten}, Karl M. and {Mizuno}, Izumi and {Mizuno}, Yosuke and {Moran}, James M. and {Moriyama}, Kotaro and {M{\"u}ller}, Cornelia and {Musoke}, Gibwa and {Mej{\'\i}as}, Alejandro Mus and {Nagar}, Neil M. and {Nakamura}, Masanori and {Narayan}, Ramesh and {Narayanan}, Gopal and {Natarajan}, Iniyan and {Neilsen}, Joey and {Neri}, Roberto and {Ni}, Chunchong and {Noutsos}, Aristeidis and {Nowak}, Michael A. and {Okino}, Hiroki and {Olivares}, H{\'e}ctor and {Ortiz-Le{\'o}n}, Gisela N. and {Oyama}, Tomoaki and {{\"O}zel}, Feryal and {Palumbo}, Daniel C.~M. and {Park}, Jongho and {Patel}, Nimesh and {Pen}, Ue-Li and {Pesce}, Dominic W. and {Pi{\'e}tu}, Vincent and {Plambeck}, Richard and {PopStefanija}, Aleksandar and {Porth}, Oliver and {P{\"o}tzl}, Felix M. and {Prather}, Ben and {Preciado-L{\'o}pez}, Jorge A. and {Psaltis}, Dimitrios and {Pu}, Hung-Yi and {Ramakrishnan}, Venkatessh and {Rao}, Ramprasad and {Rawlings}, Mark G. and {Raymond}, Alexander W. and {Rezzolla}, Luciano and {Ripperda}, Bart and {Roelofs}, Freek and {Rogers}, Alan and {Rose}, Mel and {Roshanineshat}, Arash and {Rottmann}, Helge and {Roy}, Alan L. and {Ruszczyk}, Chet and {Rygl}, Kazi L.~J. and {S{\'a}nchez}, Salvador and {S{\'a}nchez-Arguelles}, David and {Sasada}, Mahito},
        title = "{Polarimetric Properties of Event Horizon Telescope Targets from ALMA}",
      journal = {\apjl},
         year = 2021,
        month = mar,
       volume = {910},
       number = {1},
          eid = {L14},
        pages = {L14},
          doi = {10.3847/2041-8213/abee6a},
archivePrefix = {arXiv},
       eprint = {2105.02272},
 primaryClass = {astro-ph.GA},
       adsurl = {https://ui.adsabs.harvard.edu/abs/2021ApJ...910L..14G}
}

@ARTICLE{greenstein64,
       author = {{Greenstein}, Jesse L. and {Schmidt}, Maarten},
        title = "{The Quasi-Stellar Radio Sources 3C 48 and 3C 273.}",
      journal = {\apj},
         year = 1964,
        month = jul,
       volume = {140},
        pages = {1},
          doi = {10.1086/147889},
       adsurl = {https://ui.adsabs.harvard.edu/abs/1964ApJ...140....1G}
}

@ARTICLE{hazard63,
       author = {{Hazard}, C. and {Mackey}, M.~B. and {Shimmins}, A.~J.},
        title = "{Investigation of the Radio Source 3C 273 By The Method of Lunar Occultations}",
      journal = {\nat},
         year = 1963,
        month = mar,
       volume = {197},
       number = {4872},
        pages = {1037-1039},
          doi = {10.1038/1971037a0},
       adsurl = {https://ui.adsabs.harvard.edu/abs/1963Natur.197.1037H}
}

@article{hovatta12,
	Adsurl = {http://adsabs.harvard.edu/abs/2012AJ....144..105H},
	Archiveprefix = {arXiv},
	Author = {{Hovatta}, T. and {Lister}, M.~L. and {Aller}, M.~F. and {Aller}, H.~D. and {Homan}, D.~C. and {Kovalev}, Y.~Y. and {Pushkarev}, A.~B. and {Savolainen}, T.},
	Doi = {10.1088/0004-6256/144/4/105},
	Eid = {105},
	Eprint = {1205.6746},
	Journal = {\aj},
	Month = oct,
	Pages = {105},
	Primaryclass = {astro-ph.CO},
	Title = {{MOJAVE: Monitoring of Jets in Active Galactic Nuclei with VLBA Experiments. VIII. Faraday Rotation in Parsec-scale AGN Jets}},
	Volume = 144,
	Year = 2012}

@ARTICLE{hovatta19,
       author = {{Hovatta}, T. and {O'Sullivan}, S. and {Mart{\'\i}-Vidal}, I. and {Savolainen}, T. and {Tchekhovskoy}, A.},
        title = "{Magnetic field at a jet base: extreme Faraday rotation in 3C 273 revealed by ALMA}",
      journal = {\aap},
         year = 2019,
        month = mar,
       volume = {623},
          eid = {A111},
        pages = {A111},
          doi = {10.1051/0004-6361/201832587},
archivePrefix = {arXiv},
       eprint = {1803.09982},
 primaryClass = {astro-ph.GA},
       adsurl = {https://ui.adsabs.harvard.edu/abs/2019A&A...623A.111H}
}

@ARTICLE{huarte11,
       author = {{Huarte-Espinosa}, M. and {Krause}, M. and {Alexander}, P.},
        title = "{3D magnetohydrodynamic simulations of the evolution of magnetic fields in Fanaroff-Riley class II radio sources}",
      journal = {\mnras},
         year = 2011,
        month = oct,
       volume = {417},
       number = {1},
        pages = {382-399},
          doi = {10.1111/j.1365-2966.2011.19271.x},
archivePrefix = {arXiv},
       eprint = {1106.4586},
 primaryClass = {astro-ph.CO},
       adsurl = {https://ui.adsabs.harvard.edu/abs/2011MNRAS.417..382H}
}

@ARTICLE{jerrim25,
       author = {{Jerrim}, Larissa and {Shabala}, Stanislav and {Yates-Jones}, Patrick and {Krause}, Martin and {Turner}, Ross and {Stewart}, Georgia and {Power}, Chris},
        title = "{BRAiSE: Synthetic polarisation in RMHD AGN jet simulations}",
      journal = {\pasa},
         year = 2025,
        month = oct,
       volume = {42},
          eid = {e136},
        pages = {e136},
          doi = {10.1017/pasa.2025.10101},
archivePrefix = {arXiv},
       eprint = {2506.19541},
 primaryClass = {astro-ph.HE},
       adsurl = {https://ui.adsabs.harvard.edu/abs/2025PASA...42..136J}
}

@ARTICLE{jester01,
       author = {{Jester}, S. and {R{\"o}ser}, H. -J. and {Meisenheimer}, K. and {Perley}, R. and {Conway}, R.},
        title = "{HST optical spectral index map of the jet of 3C 273}",
      journal = {\aap},
         year = 2001,
        month = jul,
       volume = {373},
        pages = {447-458},
          doi = {10.1051/0004-6361:20010593},
archivePrefix = {arXiv},
       eprint = {astro-ph/0104393},
 primaryClass = {astro-ph},
       adsurl = {https://ui.adsabs.harvard.edu/abs/2001A&A...373..447J}
}

@ARTICLE{jester05,
       author = {{Jester}, S. and {R{\"o}ser}, H.-J. and {Meisenheimer}, K. and {Perley}, R.},
        title = "{The radio-ultraviolet spectral energy distribution of the jet in 3C 273}",
      journal = {\aap},
         year = 2005,
        month = feb,
       volume = {431},
        pages = {477-502},
          doi = {10.1051/0004-6361:20047021},
archivePrefix = {arXiv},
       eprint = {astro-ph/0410520},
 primaryClass = {astro-ph},
       adsurl = {https://ui.adsabs.harvard.edu/abs/2005A&A...431..477J}
}

@ARTICLE{jester07,
       author = {{Jester}, Sebastian and {Meisenheimer}, Klaus and {Martel}, Andr{\'e} R. and {Perlman}, Eric S. and {Sparks}, William B.},
        title = "{Hubble Space Telescope far-ultraviolet imaging of the jet in 3C273: a common emission component from optical to X-rays}",
      journal = {\mnras},
         year = 2007,
        month = sep,
       volume = {380},
       number = {2},
        pages = {828-834},
          doi = {10.1111/j.1365-2966.2007.12120.x},
archivePrefix = {arXiv},
       eprint = {0706.2564},
 primaryClass = {astro-ph},
       adsurl = {https://ui.adsabs.harvard.edu/abs/2007MNRAS.380..828J}
}

@misc{kameno23,
  author       = {Kameno, Seiji and Cort\'es, Paulo and Kneissl, R\¨udiger and Paladino, Rosita and Plarre, Kurt and  Radiszcz, Matias},
  title        = {AMAPOLA: ALMA Polarization Monitoring for
                   Calibrator Sources, ALMA at 10 years: Past, Present, and Future (alma2023), Puerto Varas, Chile
                  },
  month        = dec,
  year         = 2023,
  publisher    = {Zenodo},
  doi          = {10.5281/zenodo.10245760},
  url          = {https://doi.org/10.5281/zenodo.10245760},
}

@article{kass1995,
author = {Robert E. Kass and Adrian E. Raftery},
title = {Bayes Factors},
journal = {Journal of the American Statistical Association},
volume = {90},
number = {430},
pages = {773--795},
year = {1995},
publisher = {ASA Website},
doi = {10.1080/01621459.1995.10476572}
}

@ARTICLE{komugi22,
       author = {{Komugi}, Shinya and {Toba}, Yoshiki and {Matsuoka}, Yoshiki and {Saito}, Toshiki and {Yamashita}, Takuji},
        title = "{Detection of Extended Millimeter Emission in the Host Galaxy of 3C 273 and Its Implications for QSO Feedback via High Dynamic Range ALMA Imaging}",
      journal = {\apj},
         year = 2022,
        month = may,
       volume = {930},
       number = {1},
          eid = {3},
        pages = {3},
          doi = {10.3847/1538-4357/ac616e},
archivePrefix = {arXiv},
       eprint = {2203.15218},
 primaryClass = {astro-ph.GA},
       adsurl = {https://ui.adsabs.harvard.edu/abs/2022ApJ...930....3K}
}

@ARTICLE{laing80,
       author = {{Laing}, R.~A.},
        title = "{A model for the magnetic-field structure in extended radio sources.}",
      journal = {\mnras},
         year = 1980,
        month = nov,
       volume = {193},
        pages = {439-449},
          doi = {10.1093/mnras/193.3.439},
       adsurl = {https://ui.adsabs.harvard.edu/abs/1980MNRAS.193..439L}
}

@ARTICLE{li22,
       author = {{Li}, Yan-Rong and {Wang}, Jian-Min and {Songsheng}, Yu-Yang and {Zhang}, Zhi-Xiang and {Du}, Pu and {Hu}, Chen and {Xiao}, Ming},
        title = "{Spectroastrometry and Reverberation Mapping: The Mass and Geometric Distance of the Supermassive Black Hole in the Quasar 3C 273}",
      journal = {\apj},
         year = 2022,
        month = mar,
       volume = {927},
       number = {1},
          eid = {58},
        pages = {58},
          doi = {10.3847/1538-4357/ac4bcb},
archivePrefix = {arXiv},
       eprint = {2201.04470},
 primaryClass = {astro-ph.GA},
       adsurl = {https://ui.adsabs.harvard.edu/abs/2022ApJ...927...58L}
}

@ARTICLE{marchenko17,
       author = {{Marchenko}, Volodymyr and {Harris}, D.~E. and {Ostrowski}, Micha{\l} and {Stawarz}, {\L}ukasz and {Bohdan}, Artem and {Jamrozy}, Marek and {Hnatyk}, Bohdan},
        title = "{Novel Analysis of the Multiwavelength Structure of the Relativistic Jet in Quasar 3C 273}",
      journal = {\apj},
         year = 2017,
        month = jul,
       volume = {844},
       number = {1},
          eid = {11},
        pages = {11},
          doi = {10.3847/1538-4357/aa755d},
archivePrefix = {arXiv},
       eprint = {1602.01654},
 primaryClass = {astro-ph.HE},
       adsurl = {https://ui.adsabs.harvard.edu/abs/2017ApJ...844...11M}
}

@article{marti-vidal14,
	Adsurl = {http://adsabs.harvard.edu/abs/2014A%26A...563A.136M},
	Archiveprefix = {arXiv},
	Author = {{Mart{\'{\i}}-Vidal}, I. and {Vlemmings}, W.~H.~T. and {Muller}, S. and {Casey}, S.},
	Doi = {10.1051/0004-6361/201322633},
	Eid = {A136},
	Eprint = {1401.4984},
	Journal = {\aap},
	Month = mar,
	Pages = {A136},
	Primaryclass = {astro-ph.IM},
	Title = {{UVMULTIFIT: A versatile tool for fitting astronomical radio interferometric data}},
	Volume = 563,
	Year = 2014}

@ARTICLE{matthews90,
       author = {{Matthews}, A.~P. and {Scheuer}, P.~A.~G.},
        title = "{Models of Radio Galaxies with Tangled Magnetic Fields - Part Two - Numerical Simulations and Their Interpretation}",
      journal = {\mnras},
         year = 1990,
        month = feb,
       volume = {242},
        pages = {623},
          doi = {10.1093/mnras/242.4.623},
       adsurl = {https://ui.adsabs.harvard.edu/abs/1990MNRAS.242..623M}
}

@ARTICLE{meisenheimer86,
       author = {{Meisenheimer}, K. and {Heavens}, A.~F.},
        title = "{Particle acceleration in the hotspot of the jet of quasar 3C273}",
      journal = {\nat},
         year = 1986,
        month = oct,
       volume = {323},
       number = {6087},
        pages = {419-422},
          doi = {10.1038/323419a0},
       adsurl = {https://ui.adsabs.harvard.edu/abs/1986Natur.323..419M}
}

@ARTICLE{meisenheimer89,
       author = {{Meisenheimer}, K. and {Roser}, H. -J. and {Hiltner}, P.~R. and {Yates}, M.~G. and {Longair}, M.~S. and {Chini}, R. and {Perley}, R.~A.},
        title = "{The synchrotron spectra of radio hot spots.}",
      journal = {\aap},
         year = 1989,
        month = jul,
       volume = {219},
        pages = {63-86},
       adsurl = {https://ui.adsabs.harvard.edu/abs/1989A&A...219...63M}
}

@article{meyer14,
doi = {10.1088/2041-8205/780/2/L27},
url = {https://doi.org/10.1088/2041-8205/780/2/L27},
year = {2013},
month = {dec},
publisher = {The American Astronomical Society},
volume = {780},
number = {2},
pages = {L27},
author = {Meyer, Eileen T. and Georganopoulos, Markos},
title = {FERMI RULES OUT THE INVERSE COMPTON/CMB MODEL FOR THE LARGE-SCALE JET X-RAY EMISSION OF 3C 273},
journal = {The Astrophysical Journal Letters}
}

@ARTICLE{meyer16,
       author = {{Meyer}, Eileen T. and {Sparks}, William B. and {Georganopoulos}, Markos and {Anderson}, Jay and {van der Marel}, Roeland and {Biretta}, John and {Sohn}, Sangmo Tony and {Chiaberge}, Marco and {Perlman}, Eric and {Norman}, Colin},
        title = "{An HST Proper-motion Study of the Large-scale Jet of 3C273}",
      journal = {\apj},
         year = 2016,
        month = feb,
       volume = {818},
       number = {2},
          eid = {195},
        pages = {195},
          doi = {10.3847/0004-637X/818/2/195},
archivePrefix = {arXiv},
       eprint = {1601.03687},
 primaryClass = {astro-ph.GA},
       adsurl = {https://ui.adsabs.harvard.edu/abs/2016ApJ...818..195M}
}

@article{nagai16,
	Adsurl = {http://adsabs.harvard.edu/abs/2016ApJ...824..132N},
	Archiveprefix = {arXiv},
	Author = {{Nagai}, H. and {Nakanishi}, K. and {Paladino}, R. and {Hull}, C.~L.~H. and {Cortes}, P. and {Moellenbrock}, G. and {Fomalont}, E. and {Asada}, K. and {Hada}, K.},
	Doi = {10.3847/0004-637X/824/2/132},
	Eid = {132},
	Eprint = {1605.00051},
	Journal = {\apj},
	Month = jun,
	Pages = {132},
	Title = {{ALMA Science Verification Data: Millimeter Continuum Polarimetry of the Bright Radio Quasar 3C 286}},
	Volume = 824,
	Year = 2016}

@ARTICLE{neumann97,
       author = {{Neumann}, M. and {Meisenheimer}, K. and {Roeser}, H. -J.},
        title = "{Near-infrared photometry of the jet of 3C 273.}",
      journal = {\aap},
         year = 1997,
        month = oct,
       volume = {326},
        pages = {69-76},
       adsurl = {https://ui.adsabs.harvard.edu/abs/1997A&A...326...69N}
}

@article{osullivan12,
	Adsurl = {http://adsabs.harvard.edu/abs/2012MNRAS.421.3300O},
	Archiveprefix = {arXiv},
	Author = {{O'Sullivan}, S.~P. and {Brown}, S. and {Robishaw}, T. and {Schnitzeler}, D.~H.~F.~M. and {McClure-Griffiths}, N.~M. and {Feain}, I.~J. and {Taylor}, A.~R. and {Gaensler}, B.~M. and {Landecker}, T.~L. and {Harvey-Smith}, L. and {Carretti}, E.},
	Doi = {10.1111/j.1365-2966.2012.20554.x},
	Eprint = {1201.3161},
	Journal = {\mnras},
	Month = apr,
	Pages = {3300-3315},
	Title = {{Complex Faraday depth structure of active galactic nuclei as revealed by broad-band radio polarimetry}},
	Volume = 421,
	Year = 2012}

@ARTICLE{perley17,
       author = {{Perley}, R.~A. and {Meisenheimer}, K.},
        title = "{High-fidelity VLA imaging of the radio structure of 3C 273}",
      journal = {\aap},
         year = 2017,
        month = may,
       volume = {601},
          eid = {A35},
        pages = {A35},
          doi = {10.1051/0004-6361/201629704},
archivePrefix = {arXiv},
       eprint = {1609.03963},
 primaryClass = {astro-ph.GA},
       adsurl = {https://ui.adsabs.harvard.edu/abs/2017A&A...601A..35P}
}

@ARTICLE{roeser96,
       author = {{Roeser}, H. -J. and {Conway}, R.~G. and {Meisenheimer}, K.},
        title = "{The synchrotron radiation from the jet of 3C273. IV. Comparison of optical and radio morphology and polarization.}",
      journal = {\aap},
         year = 1996,
        month = oct,
       volume = {314},
        pages = {414-418},
       adsurl = {https://ui.adsabs.harvard.edu/abs/1996A&A...314..414R}
}

@ARTICLE{roeser00,
   author = {{R{\"o}ser}, H.-J. and {Meisenheimer}, K. and {Neumann}, M. and 
	{Conway}, R.~G. and {Perley}, R.~A.},
    title = "{The jet of 3C 273 observed with ROSAT HRI}",
  journal = {\aap},
   eprint = {astro-ph/0005592},
     year = 2000,
    month = aug,
   volume = 360,
    pages = {99-106},
   adsurl = {http://adsabs.harvard.edu/abs/2000A%26A...360...99R}
}

@ARTICLE{sambruna01,
       author = {{Sambruna}, Rita M. and {Urry}, C. Megan and {Tavecchio}, F. and {Maraschi}, L. and {Scarpa}, R. and {Chartas}, G. and {Muxlow}, T.},
        title = "{Chandra Observations of the X-Ray Jet of 3C 273}",
      journal = {\apjl},
         year = 2001,
        month = mar,
       volume = {549},
       number = {2},
        pages = {L161-L165},
          doi = {10.1086/319157},
archivePrefix = {arXiv},
       eprint = {astro-ph/0101299},
 primaryClass = {astro-ph},
       adsurl = {https://ui.adsabs.harvard.edu/abs/2001ApJ...549L.161S}
}

@ARTICLE{savolainen02,
       author = {{Savolainen}, T. and {Wiik}, K. and {Valtaoja}, E. and {Jorstad}, S.~G. and {Marscher}, A.~P.},
        title = "{Connections between millimetre continuum variations and VLBI structure in 27 AGN}",
      journal = {\aap},
         year = 2002,
        month = nov,
       volume = {394},
        pages = {851-861},
          doi = {10.1051/0004-6361:20021236},
       adsurl = {https://ui.adsabs.harvard.edu/abs/2002A&A...394..851S}
}

@ARTICLE{schmidt1963,
       author = {{Schmidt}, M.},
        title = "{3C 273 : A Star-Like Object with Large Red-Shift}",
      journal = {\nat},
         year = 1963,
        month = mar,
       volume = {197},
       number = {4872},
        pages = {1040},
          doi = {10.1038/1971040a0},
       adsurl = {https://ui.adsabs.harvard.edu/abs/1963Natur.197.1040S}
}

@ARTICLE{schmidt78,
       author = {{Schmidt}, G.~D. and {Peterson}, B.~M. and {Beaver}, E.~A.},
        title = "{Imaging polarimetry of the jet of M87 and 3C 273.}",
      journal = {\apjl},
         year = 1978,
        month = mar,
       volume = {220},
        pages = {L31-L36},
          doi = {10.1086/182631},
       adsurl = {https://ui.adsabs.harvard.edu/abs/1978ApJ...220L..31S}
}

@ARTICLE{stawarz02,
       author = {{Stawarz}, {\L}. and {Ostrowski}, M.},
        title = "{Radiation from the Relativistic Jet: A Role of the Shear Boundary Layer}",
      journal = {\apj},
         year = 2002,
        month = oct,
       volume = {578},
       number = {2},
        pages = {763-774},
          doi = {10.1086/342649},
archivePrefix = {arXiv},
       eprint = {astro-ph/0203040},
 primaryClass = {astro-ph},
       adsurl = {https://ui.adsabs.harvard.edu/abs/2002ApJ...578..763S}
}

@ARTICLE{tavecchio25,
       author = {{Tavecchio}, Fabrizio},
        title = "{The TeV emission of 3C273: Inverse Compton radiation from shear-accelerated high-energy electrons in the large-scale jet?}",
      journal = {\aap},
         year = 2025,
        month = dec,
       volume = {704},
          eid = {L10},
        pages = {L10},
          doi = {10.1051/0004-6361/202557289},
archivePrefix = {arXiv},
       eprint = {2511.04433},
 primaryClass = {astro-ph.HE},
       adsurl = {https://ui.adsabs.harvard.edu/abs/2025A&A...704L..10T}
}

@ARTICLE{uchiyama06,
       author = {{Uchiyama}, Yasunobu and {Urry}, C. Megan and {Cheung}, C.~C. and {Jester}, Sebastian and {Van Duyne}, Jeffrey and {Coppi}, Paolo and {Sambruna}, Rita M. and {Takahashi}, Tadayuki and {Tavecchio}, Fabrizio and {Maraschi}, Laura},
        title = "{Shedding New Light on the 3C 273 Jet with the Spitzer Space Telescope}",
      journal = {\apj},
         year = 2006,
        month = sep,
       volume = {648},
       number = {2},
        pages = {910-921},
          doi = {10.1086/505964},
archivePrefix = {arXiv},
       eprint = {astro-ph/0605530},
 primaryClass = {astro-ph},
       adsurl = {https://ui.adsabs.harvard.edu/abs/2006ApJ...648..910U}
}

@ARTICLE{willingale81,
       author = {{Willingale}, R.},
        title = "{Use of the maximum entropy method in X-ray astronomy}",
      journal = {\mnras},
         year = 1981,
        month = feb,
       volume = {194},
        pages = {359-364},
          doi = {10.1093/mnras/194.2.359},
       adsurl = {https://ui.adsabs.harvard.edu/abs/1981MNRAS.194..359W}
}

\begin{appendix}

\section{Model definitions for QU-fitting}\label{app:model_defs}
The functional forms of the four Faraday-rotation models considered in Sect.~\ref{section:core} are given here.
For each model we also list the associated parameter vector, $\boldsymbol{\theta},$ that is sampled during the QU-fitting \citep[e.g.,][]{osullivan12}.

\paragraph{(S) Single Faraday-thin}
\begin{equation}\label{eq:m1_model}
p(\lambda^2) =
p_0\,\exp\!\Big[\,2i\big(\psi_0 + RM\,\lambda^2\big)\Big].
\end{equation}
The parameter vector is
\[
\boldsymbol{\theta}_{\mathrm{S}}
= \{\,p_0,\ \psi_0,\ RM\,\}.
\]

\paragraph{(T) Single inverse-depolarization}
\begin{equation}\label{eq:m5_model}
p(\lambda^2) =
p_0\,\exp\!\Big[\,2i\big(\psi_0 + RM\,\lambda^2\big)\Big]\,
\sinc\!\Big(\frac{\Theta - RM_{\mathrm{int}}\lambda^2}{\pi}\Big),
\end{equation}
where $\Theta$ (deg) and $RM_{\mathrm{int}}$ (rad\,m$^{-2}$) describe internal differential rotation.
The parameter vector is
\[
\boldsymbol{\theta}_{\mathrm{T}}
= \{\,p_0,\ \psi_0,\ RM,\ \Theta,\ RM_{\mathrm{int}}\,\}.
\]

\paragraph{(S2) Two Faraday-thin components}
\begin{equation}\label{eq:m11_model}
p(\lambda^2) =
\sum_{k=1}^{2}
p_{0,k}\,\exp\!\Big[\,2i\big(\psi_{0,k} + RM_k\,\lambda^2\big)\Big].
\end{equation}
The parameter vector is
\[
\boldsymbol{\theta}_{\mathrm{S2}}
= \{\,p_{0,1},\ \psi_{0,1},\ RM_1,\ 
      p_{0,2},\ \psi_{0,2},\ RM_2\,\}.
\]

\paragraph{(T2) Two components with external Faraday dispersion}
\begin{equation}\label{eq:m4_model}
p(\lambda^2) =
\sum_{k=1}^{2}
p_{0,k}\,
\exp\!\Big[\,2i\big(\psi_{0,k} + RM_k\,\lambda^2\big)\Big]\,
\exp\!\big[-2\sigma_{\mathrm{RM},k}^{2}\lambda^{4}\big],
\end{equation}
with $\sigma_{\mathrm{RM},k}\ge0$ describing external Faraday dispersion.
The parameter vector is
\[
\boldsymbol{\theta}_{\mathrm{T2}}
=
\{\,p_{0,1},\ \psi_{0,1},\ RM_1,\ \sigma_{\mathrm{RM},1},\ 
      p_{0,2},\ \psi_{0,2},\ RM_2,\ \sigma_{\mathrm{RM},2}\,\}.
\]

\section{Fitted parameters}\label{app:params}
The individual parameter values for the best-fitting T2 model are given Table~\ref{table:T2param} and the corner plot is shown in Fig.~\ref{fig:m4_corner}.

\begin{figure*}
\centering
\includegraphics[width=\textwidth]{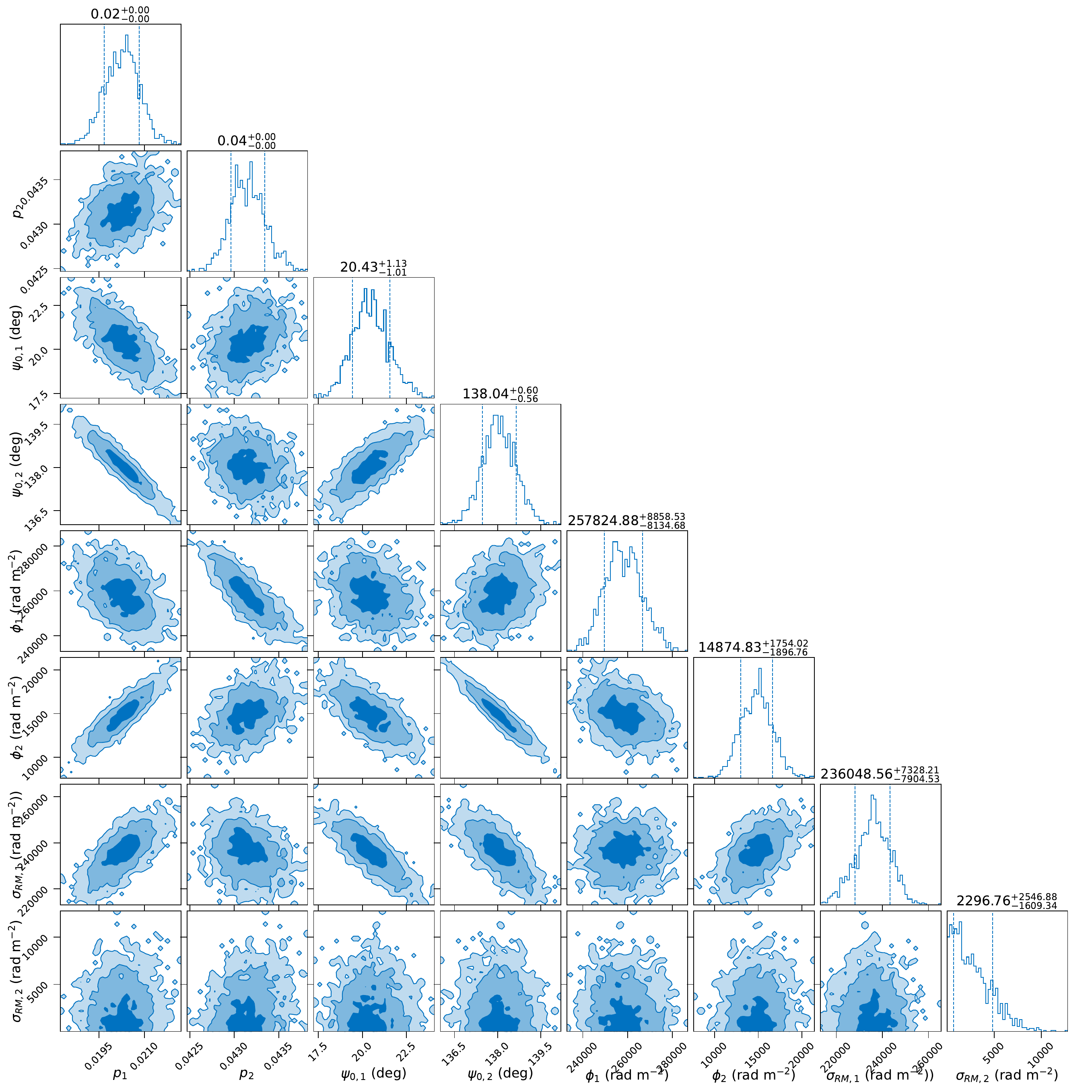}
\caption{Posterior corner plot for the best-fit two-component external-dispersion model (T2; Eq.~\ref{eq:m4_model}). 
Diagonal panels show 1D marginalized posteriors; off-diagonals show 2D covariances between parameters
(e.g., $p_{0,k}$, $\psi_{0,k}$, $RM_k$, and $\sigma_{\mathrm{RM},k}$ for $k{=}1,2$). 
Contours mark the highest posterior density credible regions (68, 90, and 95\%).}
\label{fig:m4_corner}
\end{figure*}

The posterior summaries (medians and credible intervals) and $\ln\mathcal{Z}$ for the other tested models 
are included below. 

\begin{table}
\centering
\caption{Best-fitting single Faraday-thin (S) model parameters, with reduced-$\chi^2$ ($\chi^2_r$) values and Bayesian evidence.}
\begin{tabular}{lcc}
\hline
Parameter & Value \\
\midrule
$p$ (\%) & $4.19^{+0.01}_{-0.01}$ \\
$\psi_0$ (deg) & $137.53^{+0.47}_{-0.46}$ \\
RM (rad m$^{-2}$) & $10914.92^{+1828.48}_{-1884.85}$ \\
\midrule
$\chi^2_r$ & \multicolumn{1}{c}{23.17} \\
BIC         & \multicolumn{1}{c}{-4295.08} \\
$\ln Z$ (evidence) & \multicolumn{1}{c}{2134.24} \\
\bottomrule
\end{tabular}
\end{table}

\begin{table} \centering
 \caption{Two Faraday-thin components (S2) best-fitting model parameters with reduced-$\chi^2$ ($\chi^2_r$) values and Bayesian evidence.}
\begin{tabular}{lcc}
\toprule
Parameter & Value \\
\midrule
$p_1$ (\%) & $0.79^{+0.01}_{-0.01}$ \\
$p_2$ (\%) & $3.63^{+0.01}_{-0.01}$ \\
$\psi_{0,1}$ (deg) & $33.44^{+1.09}_{-1.07}$ \\
$\psi_{0,2}$ (deg) & $147.07^{+0.24}_{-0.24}$ \\
$\phi_1$ (rad m$^{-2}$) & $350935.13^{+7529.80}_{-7411.49}$ \\
$\phi_2$ (rad m$^{-2}$) & $-12573.96^{+1631.29}_{-1604.46}$ \\
\midrule
$\chi^2_r$ & \multicolumn{1}{c}{0.63} \\
BIC         & \multicolumn{1}{c}{-12628.19} \\
$\ln Z$ (evidence) & \multicolumn{1}{c}{6296.06} \\
\bottomrule
\end{tabular}
\end{table}

\begin{table} \centering
 \caption{Single inverse-depolarization (T\textsubscript{inv}) best-fitting model parameters with reduced-$\chi^2$ ($\chi^2_r$) values and Bayesian evidence.}
\begin{tabular}{lcc}
\toprule
Parameter & Value  \\
\midrule
$p$ (\%) & $18^{+5}_{-4}$ \\
$\psi_0$ (deg) & $147.14^{+0.13}_{-0.12}$ \\
$\Theta$ (deg) & $158.34^{+4.57}_{-4.67}$ \\
RM (rad m$^{-2}$) & $-27920.28^{+541.34}_{-541.15}$ \\
RM internal (rad m$^{-2}$) & $53732.42^{+10950.46}_{-10696.49}$ \\
\midrule
$\chi^2_r$ & \multicolumn{1}{c}{2.10} \\
BIC         & \multicolumn{1}{c}{-10588.36} \\
$\ln Z$ (evidence) & \multicolumn{1}{c}{5282.02} \\
\bottomrule
\end{tabular}
\end{table}

\end{appendix}

\end{document}